\documentclass[aps,showpacs,twocolumn,superscriptaddress,prb,floatfix, 10pt]{revtex4-1}
\usepackage{graphicx,color,epsfig,rotate}
\usepackage{amssymb,amsmath,bm}
\usepackage{dcolumn}  
\usepackage[usenames,dvipsnames]{xcolor}
\usepackage{subfigure}
\usepackage{wrapfig}
\usepackage{braket}
\usepackage{bm}
\usepackage{graphicx}
\usepackage{color}
\usepackage{dcolumn}
\usepackage{array}

\newcommand{\g} \textbf{}

\begin{document}

\title{Magnetic impurities in Kondo insulators: An application to samarium hexaboride}

\author{W. T. Fuhrman}
\affiliation{Institute for Quantum Matter at Johns Hopkins University, Baltimore, MD 21218, USA}
\author{P. Nikoli\'c}
\affiliation{Department of Physics and Astronomy,\\George Mason University, Fairfax, VA 22030, USA}
\affiliation{Institute for Quantum Matter at Johns Hopkins University, Baltimore, MD 21218, USA}

\date{\today}

\begin{abstract}

Impurities and defects in Kondo insulators can have an unusual impact on dynamics that blends with effects of intrinsic electron correlations. Such crystal imperfections are difficult to avoid, and their consequences are incompletely understood. Here we study magnetic impurities in Kondo insulators via perturbation theory of the s-d Kondo impurity model adapted to small bandgap insulators. The calculated magnetization and specific heat agree with recent thermodynamic measurements in samarium hexaboride (SmB$_6$). This qualitative agreement supports the physical picture of multi-channel Kondo screening of local moments by electrons and holes involving both intrinsic and impurity bands. Specific heat is thermally activated in zero field by Kondo screening through sub-gap impurity bands and exhibits a characteristic upturn as the temperature is decreased. In contrast, magnetization obtains a dominant quantum correction from partial screening by virtual particle-hole pairs in intrinsic bands. We point out that magnetic impurities could impact de Haas-van Alphen quantum oscillations in SmB$_6$, through the effects of Landau quantization in intrinsic bands on the Kondo screening of impurity moments.

\end{abstract}

\maketitle

\section{Introduction}

Impurities within Kondo insulators are distinct from the typical electron and hole-type impurities in semiconductors \cite{riseborough2000heavy}. A popular physical picture is that the formation of a Kondo insulating ground state is predicated on a coherent lattice of localized moments that develop singlet correlations with mobile electrons \cite{Coleman2015-book}. When impurities break translational symmetry and disturb the coherence of the ground state, they become ``Kondo holes" in the Kondo lattice. The theory of non-magnetic Kondo holes has been studied extensively, revealing a novel impurity band at dilute concentrations and a collapse of the insulating state at moderate and higher concentrations \cite{schlottmann1992impurity, schlottmann1993impurity, riseborough2003collapse}.

Experimental results on impurities and defects in Kondo insulators show an analogy to the Kondo impurity model, including a resistance minimum for dilute La doping in CePd$_3$ and impurity-driven localization in La-doped CeNiSn \cite{lawrence1996kondo, takabatake1999impurity}. In addition to non-magnetic impurities, rare earth elements with substantial magnetic moments (e.g. Gd, Eu) are common impurities in Kondo insulators \cite{Kim2013a,  fuhrman2018screened}. Their presence also disrupts the coherent Kondo insulator state, yet the experimental consequences of their magnetic degrees of freedom have largely been overlooked.

The theory of magnetic impurities in metals has a long history \cite{Anderson1961, Kondo1964, Abrikosov1965, Abrikosov1965b, Yosida1966, Duke1967a, Duke1967b, Anderson1970-Kondo, Anderson1973-Kondo, Wilson1975, Haldane1978, Andrei1980, Wiegmann1981, Andrei1983, Tsvelik1983, Okiji1983, Georges1996}. Magnetic impurities in insulators have attracted much less attention so far. Nevertheless, theoretical studies of Kondo screening in gapped systems (insulators and superconductors) have reached an important result that a Kondo singlet state does form at low temperatures, just like in metallic systems, if the gap is of the order of the Kondo temperature or smaller \cite{Ogura1993, Satori1992, Saso1992}.

The most-studied Kondo insulator, samarium hexaboride (SmB$_6$), is a strongly correlated ``heavy fermion'' material and a proposed strong topological insulator (TI) with time-reversal (TR) symmetry \cite{Dzero2010, Dzero2012, Dzero2013}. The former has been established in numerous experiments over several decades now \cite{Alekseev2010, Fuhrman2014}, while the evidence for the latter is recent and growing \cite{Zhang2013, Wolgast2013, Neupane2013, Jiang2013, Kim2013a, Xu2013, Xu2014}. As a correlated TR-invariant TI, SmB$_6$ could exhibit novel physical phenomena including an exotic bulk ground state and correlated topologically protected surface states (a 2D Dirac heavy-fermion system) \cite{Nikolic2014b, Roy2014, Efimkin2014}. Experimental evidence is mounting that surface states in SmB$_6$ are affected by interactions, either among the intrinsic degrees of freedom (e.g. mediated by a collective mode), and/or involving impurities (such as Sm vacancies, which are known to proliferate at the surface).\cite{nakajima2016one, park2016topological, arab2016effects, biswas2017suppression, wolgast2015magnetotransport} The possibility of strongly interacting surface states gives SmB$_6$ special importance among the expanding family of topological materials.
 
Several experimental studies of SmB$_6$ have recently observed puzzling dynamics consistent with metallic behaviors \cite{Flachbart2006, Phelan2014, Sebastian2015, Sebastian2018, Laurita2016} despite measurements showing that SmB$_6$ is an electric and thermal-transport DC insulator in the bulk \cite{Sera1996, Kebede1996, Xu2016, Boulanger2018}, with a spectroscopically clear gap to all excitations \cite{Alekseev1995, Bouvet1998, Gorshunov1999, Fuhrman2014, Nikolic2014c}. In particular, Corbino geometry transport measurements show unambiguously the insulating nature of the bulk \cite{Eo2019}. Measurements of de Haas-van Alphen (dHvA) effect in quantum oscillations \cite{Sebastian2015, Sebastian2018} have indicated a possible 3D bulk Fermi surface in SmB$_6$, involving quasiparticles that couple to the external magnetic field but do not transport charge; other similar measurements, however, have been interpreted as resulting from 2D surface dynamics \cite{Xiang2013, LuLi2017}. Optical conductivity \cite{Laurita2016} shows a continuum-like density of states that absorb light at sub-gap energies, but with a frequency dependence that extrapolates to a vanishing DC conductivity. On the other hand, inelastic neutron scattering has not detected any apparent magnetic spectral weight in the energy range $0.15-13\;\textrm{meV}$ below the energy of the coherent spin-exciton. The implication of this absence of scattering is that the putative low-energy degrees of freedom responsible for these dynamics must be non-magnetic, have a very small moment, or be related to impurities and defects. Their footprint is seen thermodynamically \cite{Flachbart2006, Phelan2014, fuhrman2018screened} as an up-turn in the low-temperature dependence of the linear specific heat ($C/T$) with decreasing temperature, and perhaps also by neutrons as a finite lifetime of the coherent exciton mode \cite{alekseev1997influence, fuhrman2018screened}.

The observed subgap degrees of freedom in SmB$_6$ could be a window into an exotic correlated ground state. The most obvious ground state candidate inspired by the quantum oscillations and specific heat is a gapless spin or Majorana liquid with a neutral Fermi surface \cite{Miranda1993, Baskaran2015, Erten2017, Chowdhury2018, Chowdhury2018a}. A Fermi liquid of charge-neutral spinons would not conduct DC currents, but could in principle couple to an external magnetic field in a quantum oscillations experiment. While a direct minimal coupling of neutral spinons to the electromagnetic field is not possible, any non-minimal coupling involving spinon's internal degrees of freedom could hardly account for the quantum oscillations. However, fractionalized electron partons, spinons and holons, necessarily interact via an emergent gauge field. This gauge field can provide an indirect minimal coupling of spinons to the physical electromagnetic field if the two gauge fields become correlated due to quantum fluctuations of gapped charged holons \cite{Chowdhury2018, Chowdhury2018a} or through other mechanisms \cite{Motrunich2006}. Such a physical picture is indeed promising as an explanation of several experiments, but also challenged by others. Heat transport measurements \cite{Xu2016, Boulanger2018} in SmB$_6$ seem to rule out a Fermi liquid contribution of any kind, and no hint of a spinon Fermi seas was found in low-energy neutron scattering studies (at energies below the collective mode) \cite{fuhrman2018screened}. Other proposed explanations of quantum oscillations \cite{Knolle2015, Knolle2016, Zhang2016, Pal2016, Kumar2017, Fritz2018, Kawakami2019} that attempt to circumvent a neutral Fermi surface may be at odds with some experimental results, although careful consideration may be able to reconcile relevant energy and field scales \cite{riseborough2017critical}. Surface Kondo breakdown \cite{Erten2016} as well as impurities and defects \cite{fuhrman2018screened, shen2018quantum, Harrison2018} have also been scrutinized for their impact on the dHvA oscillations.

In this paper we explore an explanation of the SmB$_6$ puzzles that are clearly related to impurities, without ruling out the prospect of an exotic ground state. Our analysis builds upon studies \cite{fuhrman2018screened, fuhrman2018diamagnetic, Valentine2016} of perplexing impurity effects in SmB$_6$, which show moment-screening and dramatic enhancement of the low-energy density of states. We argue that these experiments find an explanation in a multi-channel Kondo screening of impurity moments, which is facilitated by electrons and holes in both intrinsic and impurity bands of a small-gap insulator. Our conclusions obtain from a calculation of magnetization and specific heat in the insulating s-d Kondo model, and hence should apply to generic small-gap materials with localized magnetic impurities. We will also point out the possibility that other puzzling behaviors of SmB$_6$ are affected by the dynamics of impurity magnetic moments in a correlated Kondo insulator environment. 

\begin{figure}[t]
\includegraphics[height=2.0in]{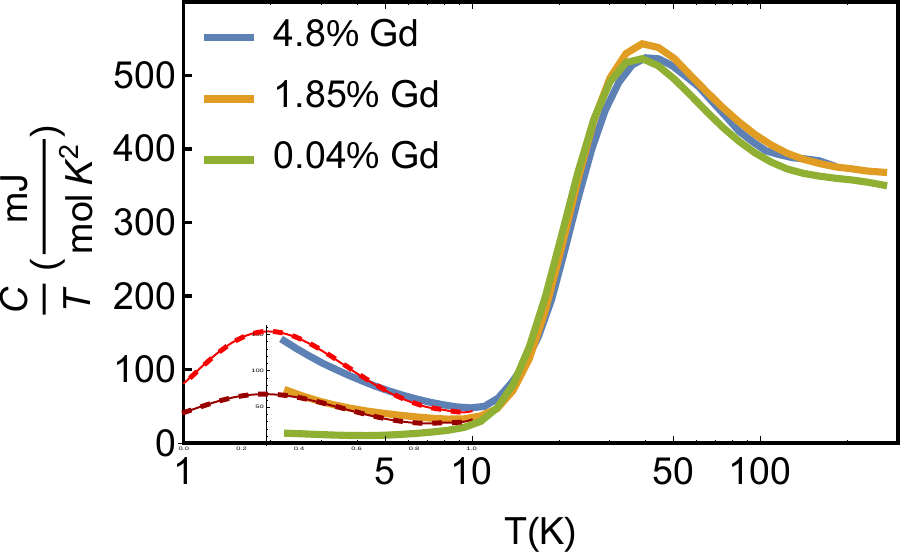}
\caption{\label{HCapFig}Specific heat of SmB$_6$ in zero magnetic field, reproduced from the supplementary material of Ref.\cite{fuhrman2018screened} (thick solid lines). The temperature $T_{\textrm{K}}^{(0)} \sim 50\textrm{ K}$ of the large peak in the data may be associated with the intrinsic Kondo temperature of the material. The analysis in this paper is concerned with the extrinsic doping-dependent specific heat upturn at $T<10\textrm{ K}$. We interpret this upturn as a result of another Kondo effect, associated with magnetic impurities in the insulating environment of SmB$_6$ (amid other possible contributions). The impurity-related Kondo temperature $T_{\textrm{K}} \ll T_{\textrm{K}}^{(0)}$ is below the temperature range of the experiment given that no low-temperature peak was reached. Two dotted thin lines are theoretical fits to the data (with subtracted $n_i=0.04\%\;\textrm{Gd}$ background) and extrapolations to lower temperatures. We used the $\beta\Delta\gg 1$ formula for $\delta c$ in Eq.\ref{Summary} and applied it near the limit of its validity $\beta\Delta\sim1$ ($\Delta\sim5.0\;\textrm{K}=0.43\;\textrm{meV}$). Even though the fits are reasonably good, this theory lacks quantitative accuracy due to a number of simplifications and approximations.}
\end{figure}

Our previous thermodynamic studies \cite{fuhrman2018screened} included measurements of magnetization and specific heat in a variety of samples with different controlled levels of impurity doping. Magnetization incorporates a background Van Vleck component related to Sm$^{2+}$, which was subtracted. The remaining magnetization shows the temperature and field dependence typical for a paramagnet of decoupled magnetic moments. We can independently extract the effective moment and concentration of impurities from the magnetization $m(\mu_0 H)$. We found that the concentration of magnetic moments was proportional to the amount of gadolinium doping, sensitive to the hundreds of ppm level. Hence, magnetization is a highly-sensitive characterization tool for a wide range of common magnetic impurities in SmB$_6$. Furthermore, the linear specific heat ($C/T$) at zero field, shown in Fig.\ref{HCapFig}, deviates from the typical insulating or even metallic behavior. It features an up-turn in its temperature dependence as the temperature is lowered well below the characteristic scale set by the SmB$_6$ gap. The amount of upturn is proportional to the amount of doping. Isolated magnetic moments due to low-density impurities in an insulator do not have capacity to store heat in zero field, so the observed specific heat must be attributed to their interaction with some additional degrees of freedom -- which are either gapless or live at very low finite energies in order to produce a seemingly non-thermally activated response. This merits our interest in an \emph{extrinsic} Kondo impurity dynamics. The intrinsic Kondo insulator physics and band topology do not seem to be important for the understanding of the impurity-related thermodynamics in SmB$_6$, and hence are not of any concern here.

\section{Summary of the analysis and conclusions}

This section describes the foundation of our analysis, specifies its validity and limitations, and states all important results. Here we provide a self-contained discussion of how the complex thermodynamic behaviors of SmB$_6$ can be theoretically understood in terms of an interaction between magnetic impurities and gapped quasiparticles. Following this section is the development of our theory. Section \ref{secI1} introduces the theoretical model, and Section \ref{secI2} reviews the thermodynamics of decoupled insulating electrons and local moments. The first-order perturbation theory is analyzed in Section \ref{secI3a}, but our main results stem from the second order perturbation theory: we separately discuss magnetization in Section \ref{secI3b} and specific heat in Section \ref{secI3c}. The lengthy details of all calculations are given in appendices. The final Section \ref{secConclusions} contains a brief summary of essential conclusions, and explores implications for the nature of quasiparticles in Kondo insulators. There we point out a physical mechanism which enables the Kondo screening of magnetic impurities to contribute dHvA effect -- possibly of some interest in the quest to understand the puzzling quantum oscillations in SmB$_6$ and YbB$_{12}$.

We begin by discussing the theory of recent magnetization and specific heat measurements in SmB$_6$. Thermodynamic experimental observations \cite{fuhrman2018screened} are consistent with a tendency of electrons in intrinsic and impurity bands to screen the localized magnetic moments introduced by rare earth impurities. Kondo impurity screening is indeed possible in gapped systems at low temperatures \cite{Ogura1993, Satori1992, Saso1992} when the Kondo temperature scale $k_{\textrm{B}}T_{\textrm{K}}$ is comparable or larger than the gap $\Delta$.

The simplest theoretical model of a Kondo insulator with magnetic impurities is the following adaptation of the s-d model's Hamiltonian:
\begin{equation}\label{sd}
H=\sum_s\left\lbrack\int d^{3}k\,E_{s{\bf k}}^{\phantom{\dagger}}\psi_{s{\bf k}}^{\dagger}\psi_{s{\bf k}}^{\phantom{\dagger}}
  -J\sum_{i=1}^{N_{i}}{\bf S}_{{\bf r}_i}^{\phantom{\dagger}}\cdot\psi_{s{\bf r}_{i}}^{\dagger}\frac{\boldsymbol{\sigma}}{2}\psi_{s{\bf r}_{i}}^{\phantom{\dagger}}
  \right\rbrack \ .
\end{equation}
The quasiparticles are described by field operators $\psi_s$ in two bands $s=\pm 1$ separated by a gap, and $N_i$ local moments scattered at locations ${\bf r}_i$ are described by spin operators ${\bf S}_{{\bf r}_i}$. This minimalistic model focuses only on the antiferromagnetic Kondo interaction $J<0$ between the magnetic impurities and quasiparticles, without seeking to capture the nature of the ground state, correlations among quasiparticles or collective modes in a Kondo insulator. The main simplification built into the model is the treatment of both quasiparticles and local moments as effective $S=\frac{1}{2}$ spin degrees of freedom with the same coupling to the external field. This reduces the technical complexity of calculations without jeopardizing the qualitative nature of conclusions. However, since magnetic impurities like gadolinium have a large moment, the price to pay is an inadequate description of underscreening that takes place in the low-temperature Kondo state \cite{Coleman2015-book}. 


We calculate magnetization up to saturating fields and specific heat in zero field using perturbation theory in the model (\ref{sd}). Our main results can be summarized by the following corrections to magnetization density $\delta m$ and zero-field specific heat $\delta c$ in a Kondo insulator (in the $\hbar=1$ units that we use throughout the paper):
\begin{eqnarray}\label{Summary}
\delta m &=&
  \begin{cases}
    -c_1 n_{i} \,\frac{(J p^3)^{2}}{\Delta}\,\beta \frac{\tanh(\beta h)}{\cosh^{2}(\beta h)} & ,\; \beta\Delta \gg 1 \\[0.05in]
   c_2 n_{i}\,Jp^3\frac{\beta}{(\beta\Delta)^{3}} \frac{\tanh(\beta h)\lbrack1+\cosh^{2}(\beta h)\rbrack}{\cosh(\beta h)} & ,\; \beta\Delta \ll 1
  \end{cases} \nonumber \\[0.08in]
\delta c &\approx&
  \begin{cases}
     c_3 n_{i}k_{\textrm{B}}\left(\frac{J p^3}{\Delta}\right)^2 (\beta\Delta)^{\frac{3}{2}}e^{-\beta\Delta} & ,\; \beta\Delta \gg 1 \\[0.05in]
     c_4 n_{i}k_{\textrm{B}}\,(\beta J p^3)^{2} & ,\; \beta\Delta \ll 1
  \end{cases}
\end{eqnarray}
These are only the dominant corrections to the response of decoupled quasiparticles and local moments. $c_{1,2,3,4}$ are positive numerical coefficients, $\beta=(k_{\textrm{B}}T)^{-1}$ is inverse temperature, $h$ is the Zeeman energy of both quasiparticle and impurity spins aligned with the external magnetic field (assumed to be the same for simplicity), $2\Delta$ is the bandgap ($\Delta\gg h$), and $n_i=N_i/V$ is the concentration of impurity moments. The formulas are limited to temperatures below a high-energy cut-off scale $W$ ($\beta W\gg 1$). A microscopic momentum scale $p$, determined from the high-energy quasiparticle spectrum, is combined with the Kondo coupling $J$ to produce an energy scale $j=J p^d$. It should be noted that $j$ is \emph{not} related to the intrinsic Kondo temperature $T_{\textrm{K}}^{(0)} \sim 50\textrm{ K}$.

The perturbation theory is controlled by the parameter $x=j/\Delta$. It contains an instability if the quasiparticles collectively form a spin-singlet with a magnetic impurity in the ground state. Therefore, the perturbation theory is valid only in conditions when such a collective screening is not developed \cite{Yosida1991}. This generally corresponds to temperatures above a Kondo scale $T_{\textrm{K}}$. In the case of SmB$_6$, the Kondo temperature $T_{\textrm{K}}$ related to magnetic impurities is lower than $1\textrm{ K}$, judging by the specific heat measured \cite{fuhrman2018screened} in SmB$_6$ and depicted in Fig.\ref{HCapFig}. Hence, our results qualitatively apply to a broad temperature regime $\Delta/k_{\textrm{B}}>T>T_{\textrm{K}}$ that probes the subgap dynamics. The intrinsic Kondo-hybridization gap $\Delta$ is well-formed near the lower end of this temperature range, so we are justified neglecting its weak residual temperature dependence and all other aspects of the intrinsic Kondo dynamics. Given $\Delta \gg k_{\textrm{B}} T_{\textrm{K}}$, the electrons in intrinsic bands are not collectively involved in the screening of the impurity moments at any temperature \cite{Ogura1993, Satori1992, Saso1992}, although local partial screening, which we calculate, does occur. We will discuss shortly the need to also consider electrons in impurity bands at much lower energies -- they appear to be responsible for the specific heat behavior, and limit the validity of perturbation theory to $T>T_{\textrm{K}}$.

The essential features of the above response functions are: (i) specific heat is thermally activated unless the Kramers degeneracy of local moments is lifted or gap closed; (ii) magnetization is not thermally activated -- it receives a quantum correction at the second order of perturbation theory by virtual particle-hole pairs that partially screen the local moments. A thermally activated component of magnetization is also found at the first order of perturbation theory, but it is not dominant at low temperatures.

The properties of the calculated $\delta c$ and $\delta m$ that are immediately consistent with the experiment \cite{fuhrman2018screened} include: (i) the system is an electric insulator, (ii) both corrections of thermodynamic responses are proportional to the impurity concentration $n_i$, (iii) magnetization is reduced in comparison to that of isolated moments (i.e. the effective moment of impurities is renormalized to a smaller value as antiferromagnetic Kondo screening with $J<0$ takes place), (iv) magnetization is not thermally activated, and (v) specific heat shows an upturn as the temperature is reduced both in the high $\beta\Delta<1$ and low $\beta\Delta>1$ temperature regimes. However, difficulties arise with attempts to fully understand specific heat: an upturn in some samples is experimentally seen down to millikelvin temperatures. This can be reconciled with the present model only if the quasiparticle spectrum features an extremely small gap, much smaller than the intrinsic $\sim\Delta$ gap of SmB$_6$.

\begin{figure}[t]
\includegraphics[height=2.2in]{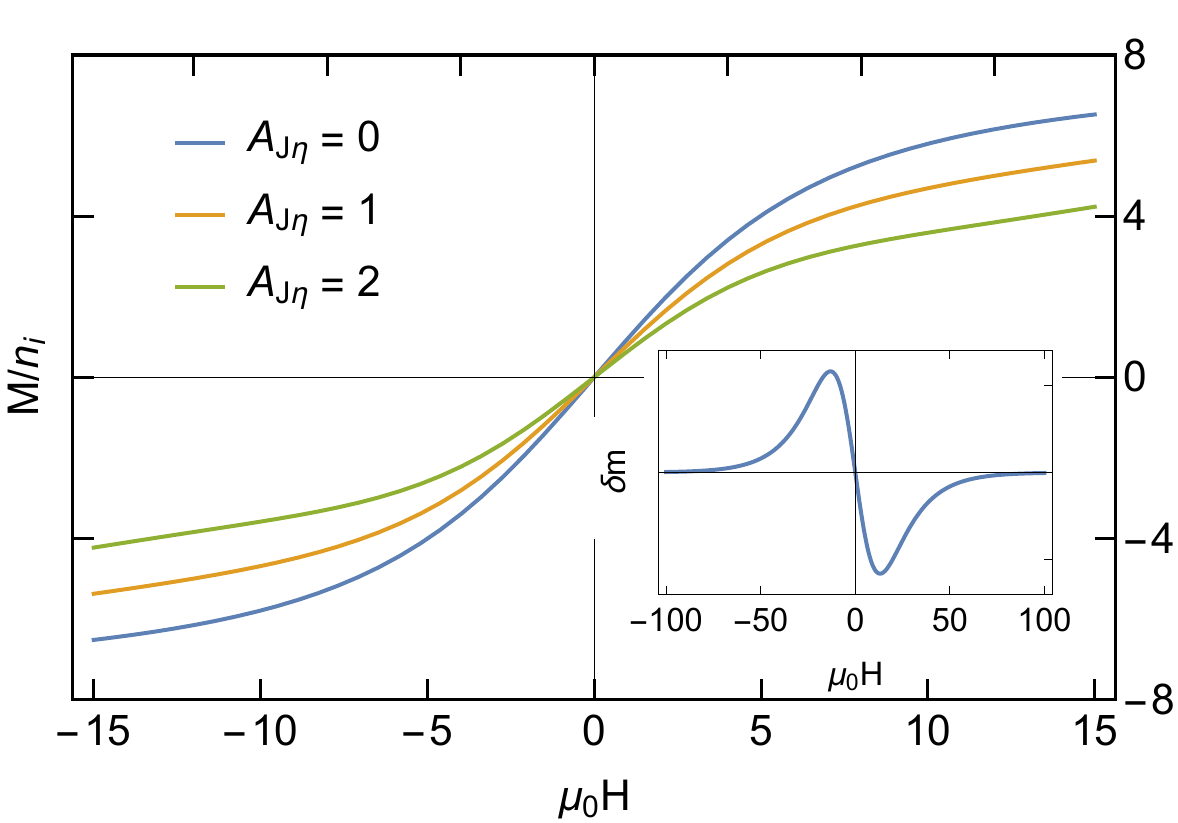}
\caption{\label{MagFig} 
Impurity magnetization of moments with $8\mu_B$ at 10 K (e.g. as for Gd \cite{fuhrman2018screened}). The amplitude of the correction is modified by impurity concentration, especially through changes in the gap scale, known to be sensitive to impurities and defects.\cite{Valentine2016} $A_J\eta$ encompasses the constant prefactor in Eq.~\ref{Summary}. Inset shows that at high fields the full unscreened impurity moment is recovered (estimated from Eg.\ref{Summary} with an adjustment for the actual large magnetic moment of Gd).}
\end{figure}

In order to resolve the problem of having an insulating transport behavior with an apparent presence of screened extrinsic magnetic moments in SmB$_6$, we suggest that multiple insulating Kondo channels give rise to the observed thermodynamics. Optical conductivity \cite{Laurita2016} provides evidence of a density of states that spans the sub-gap range of energies. This has been explored theoretically in the ``Kondo hole'' picture, when an in-gap impurity band locks the Fermi-level or comes with lower-energy localized magnetic excitations \cite{schlottmann1993impurity}. Micro-gaps $\Delta_i$ can develop as the impurity bands form and create a new channel for Kondo screening that appears not thermally activated in the specific heat measurements \cite{fuhrman2018screened}. Our calculations access this Kondo channel in its ``high temperature'' regime $\beta\Delta_i<1$. Variability in this temperature range of the heat capacity is clearly related to impurities and defects, and previous analysis of heat capacity on other samples has included Schottky anomalies\cite{Phelan2014, allen1980mixed} to partially account for the upturn in linear heat capacity. At the same time, magnetization can be contributed both by the impurity and the intrinsic electron-hole channels, since the latter is not thermally activated. Hence, the calculated response functions exhibit all essential features of their measured counterparts in the experiment \cite{fuhrman2018screened} (see Fig.\ref{MagFig}).

It will become apparent later that the momentum scale $p$ is related to the gap $\Delta$, cut-off energy $W$ and the average density of states $\rho$ in the quasiparticle bands associated with a Kondo channel:
\begin{equation}
p^{3} \sim \rho W\left(\frac{\Delta}{W}\right)^{3} \ .
\end{equation}
Therefore, if we compute from (\ref{Summary}) the ratio of the dominant magnetization correction magnitudes in the intrinsic ($\Delta_0$) and impurity ($\Delta_i$) Kondo channels:
\begin{eqnarray}
\frac{\delta m_{2}^{(0)}}{\delta m_{1}^{(i)}} &\sim& \frac{(Jp_{0}^{3})^{2}/\Delta_{0}}{Jp_{i}^{3}}(\beta\Delta_{i})^{3}
   \xrightarrow{\beta\Delta_{i}\sim1}\frac{(Jp_{0}^{3})^{2}/\Delta_{0}}{Jp_{i}^{3}} \nonumber \\
&\sim& \frac{J\rho_{0}W_{0}}{\Delta_{0}}\times\frac{\rho_{0}}{\rho_{i}}\left(\frac{\Delta_{0}}{\Delta_{i}}\right)^{3}\left(\frac{\Delta_{0}}{W_{0}}\right)^{3}
   \left(\frac{W_{i}}{W_{0}}\right)^{2} \ ,
\end{eqnarray}
we can find a natural possibility realized with $\Delta_0 \gg \Delta_i$ and $\rho_0 \gg \rho_i$ that the quantum contribution of the intrinsic channel is notably larger than the thermal contribution of the impurity channel (even in the perturbative limit $J\rho_{0}W_{0}/\Delta_{0} \ll 1$). Note that the energy cut-offs $W$ are limited both by the bandwidths and microscopic properties of the Kondo interaction (e.g. spatial range), so is it not unnatural to have comparable scales $W_0 \sim \Delta_0$, and even $W_i \sim W_0$ when impurity levels fill up the gap.

In simple words, the thermodynamic experiment \cite{fuhrman2018screened} may be revealing a thermal correction to specific heat in the impurity Kondo channel and a quantum correction to magnetization in the intrinsic Kondo channel. Both are determined at the second order of perturbation theory and proportional to $J^2$ when the quasiparticles are gapped. This interpretation is of particular importance because the coefficient of the specific heat now matches that of the correction to magnetization in the scaling found empirically in our previous experiment\cite{fuhrman2018screened}. This is a distinct contrast to the metallic s-d model, where corrections to specific heat are $\propto (J\eta)^4$ and magnetization corrections are $\propto J\eta$, with $\eta$ being the density of states at the Fermi energy. Given that the scaling was consistent over more than two orders of magnitude of impurity concentration, this insulating model represents a substantial improvement over a direct comparison to the metallic Kondo impurity effect for the case of SmB$_6$.

\section{Perturbation theory of an Insulating Kondo impurity model}

Here we analyze thermodynamics of an s-d model of Kondo impurities in an insulator, using perturbation theory. We calculate magnetization in an external magnetic field up to saturation, and specific heat in zero field. It turns out that magnetization corrections to the response of isolated local moments are dominated by a quantum process at the second order of perturbation theory in which virtual particle-hole pairs screen the local moments via Kondo coupling. In contrast, the zero-field specific heat is thermally activated but shaped by processes that also start at the second order of perturbation theory. These results provide foundation for the physical picture we build -- and conclusion that Kondo-like impurities likely play a significant role in some metallic-looking behaviors of SmB$_6$.

\subsection{Model}\label{secI1}

The s-d model we study is given by the Hamiltonian:
\begin{equation}
H_{\textrm{sd}}=\int \frac{d^{d}k}{(2\pi)^d}\,\Psi_{{\bf k}}^{\dagger}h_0^{\phantom{\dagger}}\Psi_{{\bf k}}^{\phantom{\dagger}}
  -J\sum_{i=1}^{N_{i}}{\bf S}_{{\bf r}_{i}}^{\phantom{\dagger}}\cdot
  \Psi_{{\bf r}_{i}}^{\dagger}\frac{1\otimes\boldsymbol{\sigma}}{2}\Psi_{{\bf r}_{i}}^{\phantom{\dagger}} \ .
\end{equation}
It describes a band insulator of electrons and localized magnetic moments in $d$ dimensions coupled by the Kondo term ($J$). We use a simple band-insulator energy spectrum
\begin{equation}
E_{s{\bf k}}^{\phantom{2}}=s\sqrt{\epsilon_{{\bf k}}^{2}+\Delta^{2}}-\mu
\end{equation}
with a band index $s=\pm 1$ and bandgap $2\Delta$, obtained from a non-interacting two-orbital Hamiltonian:
\begin{equation}
h_{0}=\left(\begin{array}{cccc}
\epsilon_{{\bf k}}-\mu & 0 & \Delta & 0\\
0 & \epsilon_{{\bf k}}-\mu & 0 & \Delta\\
\Delta & 0 & -\epsilon_{{\bf k}}-\mu & 0\\
0 & \Delta & 0 & -\epsilon_{{\bf k}}-\mu
\end{array}\right) \ .
\end{equation}
This representation is compatible with spinor field operators $\Psi$ whose components $\psi_{n\alpha}$ are labeled by an orbital index $n\in\lbrace 1,2 \rbrace$ and spin $\alpha$:
\begin{equation}
\Psi=\left(\begin{array}{c}
\psi_{1\uparrow}\\
\psi_{1\downarrow}\\
\psi_{2\uparrow}\\
\psi_{2\downarrow}
\end{array}\right) \ .
\end{equation}
For simplicity, we work with $\epsilon_{\bf k} = v |{\bf k}|$ that makes the momentum dependence $E_{s{\bf k}}$ formally relativistic at high energies; this microscopic feature is ultimately collected into a single momentum scale and otherwise not essential for our conclusions.

Local moments sit at randomly scattered positions ${\bf r}_i$ and have an average concentration $n_i = N_i/V$ in the system of volume $V$. We consider spin $S=\frac{1}{2}$ local moments and represent their spin operators
\begin{equation}
{\bf S}_{{\bf r}_i}^{\phantom{x}} = z_{{\bf r}_i}^\dagger \boldsymbol{\sigma} z_{{\bf r}_i}^{\phantom{\dagger}}
\end{equation}
in terms of two-component field operators $z^\dagger, z$ for electrons localized at impurity sites ($\boldsymbol{\sigma}$ is the vector of Pauli matrices). We assume that the moments are too far apart to interact with one another.

We calculate magnetization density $m(h,T)$ and specific heat $c(h,T)$ as functions of the applied magnetic field $h$ and temperature $T$:
\begin{equation}\label{ThermoDyn}
m=-\frac{\partial g}{\partial h} \quad,\quad s=-\frac{\partial g}{\partial T}\quad,\quad c=T\frac{\partial s}{\partial T} \ ,
\end{equation}
from the free energy density $g$:
\begin{equation}\label{sdF}
g = -\frac{k_{\textrm{B}}T}{V}\log(\Xi) \ .
\end{equation}
The partition function $\Xi$ is obtained from the imaginary-time path-integral in grand canonical ensemble, with chemical potentials $\mu$ for mobile electrons and $-i\lambda$ for impurity electrons:
\begin{eqnarray}\label{pathint}
\Xi &=& \int\mathcal{D}z\mathcal{D}z^{\dagger}\mathcal{D}\psi\mathcal{D}\psi^{\dagger}
  \exp \Biggl\lbrace -\int\limits_0^\beta \!\! d\tau \biggl\lbrack \\ \nonumber
&& \!\!\!\!\!\!\!\! \sum_{s} \int \frac{d^{d}k}{(2\pi)^d}\; \psi_{s{\bf k}}^{\dagger}\left(\frac{\partial}{\partial\tau} 
   + E_{s{\bf k}}^{\phantom{2}}-\mu-h\sigma^{z}\right) \psi_{s{\bf k}}^{\phantom{\dagger}} \nonumber \\
&& \!\!\!\!\!\!\!\! -J\sum_{i=1}^{N_{i}}{\bf S}_{{\bf r}_{i}}^{\phantom{\dagger}}\!\sum_{ss'}
   \!\int\!\!\frac{d^{d}k}{(2\pi)^{d}}\frac{d^{d}k'}{(2\pi)^{d}}e^{i({\bf k}'-{\bf k}){{\bf r}_i}}
     U_{s{\bf k},s'{\bf k}'}^{\phantom{\dagger}} \psi_{s{\bf k}}^{\dagger}\frac{\boldsymbol{\sigma}}{2}\psi_{s'{\bf k}'}^{\phantom{\dagger}} \nonumber \\
&& \!\!\!\!\!\!\!\! +\sum_{i}\left( z_{i}^{\dagger}\frac{\partial z_{i}}{\partial\tau}-hz_{i}^{\dagger}\sigma^{z}z_{i}^{\phantom{\dagger}}
   +i\lambda z_{i}^{\dagger}z_{i}^{\phantom{\dagger}}\right) \biggr\rbrack \Biggr\rbrace \ , \nonumber
\end{eqnarray}
where $\beta = (k_{\textrm{B}} T)^{-1}$ and $k_{\textrm{B}}$ is Boltzmann constant. For simplicity, we assume that mobile and localized electrons couple the same way to the magnetic field $h$. Representing the Kondo coupling in the band basis, with two-component band spinors $\psi_{s{\bf k}}$, requires the following vertex function:
\begin{widetext}
\begin{equation}\label{Usk}
U_{s{\bf k},s'{\bf k}'}=\frac{\Delta^{2}+\Bigl(s\sqrt{\epsilon_{{\bf k}}^{2}+\Delta^{2}}-\epsilon_{{\bf k}}^{\phantom{2}}\Bigr)\Bigl(s'\sqrt{\epsilon_{{\bf k}'}^{2}+\Delta^{2}}-\epsilon_{{\bf k}'}^{\phantom{2}}\Bigr)}{2\sqrt{\left(\Delta^{2}+\epsilon_{{\bf k}}^{2}-\epsilon_{{\bf k}}^{\phantom{2}}s\sqrt{\epsilon_{{\bf k}}^{2}+\Delta^{2}}\right)\left(\Delta^{2}+\epsilon_{{\bf k}'}^{2}-\epsilon_{{\bf k}'}^{\phantom{2}}s'\sqrt{\epsilon_{{\bf k}'}^{2}+\Delta^{2}}\right)}}\xrightarrow{\Delta\to0\;\vee\;{\bf k}'={\bf k}}\delta_{ss'}
\end{equation}
\end{widetext}

Using a spinor $z$ to generate the quantum dynamics of local moments has the crucial advantage of being amenable to Wick's theorem in perturbation theory. However, unphysical states with unoccupied and double-occupied impurity sites are also generated. Popov and Fedotov have shown \cite{Popov1988} that these unphysical states can be completely eliminated from the partition function of an arbitrary interacting theory simply by setting the chemical potential of localized electrons to $i\lambda = i\pi/2\beta$, without an adverse effect on physical states. We apply this trick in all final formulas to faithfully deduce the dynamics of local moments. It should be also noted that the constructed spectrum has no energy bounds, so we must introduce an energy cut-off $W$ (bandwidth) and regularize the field theory in order to not predict an infinite degeneracy pressure. The latter amounts to adding a constant term to the action, proportional to the volume $V$, which cancels the unphysical contributions to pressure -- we do not explicitly show this procedure.

\subsection{Unperturbed free electrons and local moments}\label{secI2}

We proceed by calculating $\Xi$ first at the zeroth order of perturbation theory $J=0$. In this case, $\Xi = \Xi_{\textrm{e}} \Xi_{\textrm{m}}$ factorizes into the textbook expressions for the grand canonical partition functions of free ``conduction'' electrons (c) and local moments (m):
\begin{eqnarray}\label{bareXi}
\log(\Xi_{\textrm{c}})&=&VA_d\,e^{-\beta\Delta}\cosh(\beta\mu)\cosh(\beta h) \nonumber \\
\log(\Xi_{\textrm{m}})&=&N_{i}\log\Bigl\lbrack 2\cosh(\beta h)\Bigr\rbrack \ ,
\end{eqnarray}
where:
\begin{equation}
A_d = \frac{4S_{d}\Gamma\left(\frac{d}{2}+1\right)(2\beta\Delta)^{d/2}}{d(2\pi\beta v)^{d}} \quad,\quad S_{d}=\frac{2\pi^{d/2}}{\Gamma\left(\frac{d}{2}\right)}
\end{equation}
and $\Gamma$ is Gamma function. Magnetization density $m$ and specific heat $c$ of electrons in a band-insulator are thermally activated:
\begin{eqnarray}\label{sdiZero}
m_{\textrm{c}} &=& A_d\,e^{-\beta\Delta}\cosh(\beta\mu)\sinh(\beta h) \\
s_{\textrm{c}} &=& k_{\textrm{B}}A_d\,e^{-\beta\Delta}(\beta\Delta)\cosh(\beta\mu)\cosh(\beta h) \nonumber \\
c_{\textrm{c}} &=& k_{\textrm{B}}A_d\,e^{-\beta\Delta}(\beta\Delta)^{2}\cosh(\beta\mu)\cosh(\beta h) \nonumber \ .
\end{eqnarray}
Note that $\mu=0$ corresponds to the Fermi energy sitting at the middle of the band-gap, and the field dependence is meaningful only in small fields $h\ll\Delta$. The contribution of decoupled local moments with concentration $n_i$ is:
\begin{eqnarray}\label{sdZero}
m_{\textrm{m}} &=& n_{i}\tanh(\beta h) \\
c_{\textrm{m}} &=& k_{\textrm{B}}n_{i}\frac{(\beta h)^{2}}{\cosh^{2}(\beta h)} \nonumber
\end{eqnarray}
at any temperature and magnetic field. The magnetization of local moments exhibits a linear dependence on small magnetic fields $\beta h\ll 1$ and saturates in large magnetic fields $\beta h \gg 1$. The same overall behavior of the measured magnetization in doped SmB$_6$, proportional to the doping concentration $n_i$, provides evidence that the doped impurities carry magnetic moments. However, the isolated magnetic moments have no heat capacity in the absence of magnetic field ($h=0$), which is where an excess specific heat is observed in the experiment. This means that the doped local moments in SmB$_6$ must be coupled to additional degrees of freedom. We discuss this coupling next.

\subsection{Perturbation theory}\label{secI3}

The perturbative expansion of the free energy (\ref{sdF}) is the sum of connected vacuum Feynman diagrams:
\begin{equation}\label{sdFa}
\log(\Xi)=\log(\Xi_{\textrm{c}})+\log(\Xi_{\textrm{m}})+\sum_{n=1}^{\infty}F_{n}
\end{equation}
where $\Xi_{\textrm{c}}$ and $\Xi_{\textrm{m}}$ are given by (\ref{bareXi}) and $F_n$ is the sum of $n^{\textrm{th}}$ order diagrams. The bare propagators $G$ of ``conduction'' electrons and $D$ of local moments are given by matrices operating in the two-component spinor space:
\begin{eqnarray}\label{sdGreen}
G(s,{\bf k},\omega_{n}) &=& \frac{1}{i\omega_{n}-(E_{s{\bf k}}-\mu)+h\sigma^{z}} \\
D^{ij}(\Omega_{n}) &=& \frac{\delta_{ij}}{i\Omega_{n}-i\lambda+h\sigma^{z}} \nonumber
\end{eqnarray}
$i,j=1,\dots,N_i$ enumerate impurity sites, and $\omega_n, \Omega_n$ are Fermionic Matsubara frequencies that take values $\omega_n = (2n+1)\pi\times k_{\textrm{B}}T$ for integer $n$. The matrix elements of these propagators, indexed by $\alpha,\beta=\pm 1$ spin-projection states along the $\hat{\bf z}$ axis are:
\begin{eqnarray}\label{sdGD}
G_{\alpha\alpha'}(s,{\bf k},\omega_{n}) &=& \frac{1}{2} \sum_{\sigma=\pm 1} 
  \frac{\delta_{\alpha\alpha'}^{\phantom{z}}+\sigma\sigma_{\alpha\alpha'}^{z}}{i\omega_{n}-(E_{s{\bf k}}-\mu-h\sigma)} \nonumber \\
D^{ij}_{\beta\beta'}(\Omega_{n}) &=& \frac{\delta_{ij}}{2} \sum_{\sigma=\pm 1} 
  \frac{\delta_{\beta\beta'}^{\phantom{z}}+\sigma\sigma_{\beta\beta'}^{z}}{i\Omega_{n}-i\lambda+h\sigma} \ .
\end{eqnarray}
The bare vertex for the Kondo coupling at $\omega+\Omega=\omega'+\Omega'$ is:
\begin{eqnarray}\label{sdVertex}
&& V_{\alpha\alpha'\beta\beta'}(\omega,s,{\bf k}\; ; \;\omega',s',{\bf k'}\; ; \;i,\Omega\; ; \;j,\Omega') = \\
&& \quad = \frac{J}{2\beta}\delta_{ij} e^{i({\bf k}-{\bf k}'){\bf r}_{i}} \boldsymbol{\sigma}_{\alpha\alpha'}\boldsymbol{\sigma}_{\beta\beta'}
  U_{s{\bf k},s'{\bf k}'} \nonumber \\
&& \quad = \frac{J}{2\beta}\delta_{ij} e^{i({\bf k}-{\bf k}'){\bf r}_{i}}
  (2\delta_{\alpha\beta'}\delta_{\beta\alpha'}-\delta_{\alpha\alpha'}\delta_{\beta\beta'})U_{s{\bf k},s'{\bf k}'} \nonumber
\end{eqnarray}
with $U_{s{\bf k},s'{\bf k}'}$ given by (\ref{Usk}).

\begin{figure}
\subfigure[{}]{\epsfig{file = 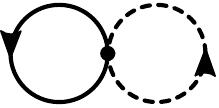, height=0.35in}}
\subfigure[{}]{\epsfig{file = 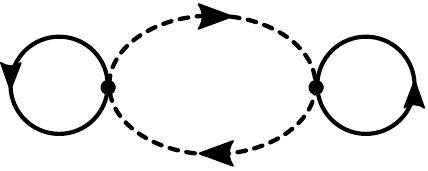, height=0.35in}}
\subfigure[{}]{\epsfig{file = 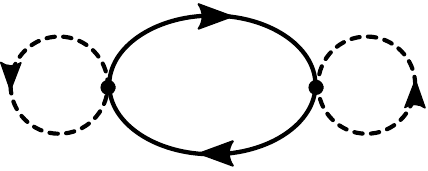, height=0.35in}}
\subfigure[{}]{\epsfig{file = 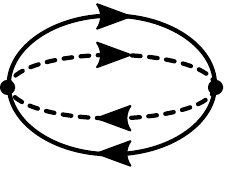, height=0.35in}}
\caption{\label{vacdiag}The connected vacuum Feynman diagrams that can contribute to free energy up to the second order of perturbation theory. Solid lines represent conduction electron propagators, and dashed lines represent impurity propagators.}
\end{figure}

\subsubsection{First order corrections}\label{secI3a}

The first-order connected vacuum diagram shown in Fig.\ref{vacdiag}(a) is:
\begin{eqnarray}\label{sdF1}
&& \!\! F_1 = (-1)^{2} \frac{J}{2\beta}(2\delta_{\alpha\beta'}\delta_{\beta\alpha'}-\delta_{\alpha\alpha'}\delta_{\beta\beta'}) \\
&& \times \sum_{i}\sum_{\omega_n\Omega_n}\sum_s\int\frac{d^{d}k}{(2\pi)^{d}}\;G^{\phantom{x}}_{\alpha\alpha'}(s,{\bf k},\omega_n)D^{ii}_{\beta\beta'}(\Omega_n)
   U_{s{\bf k},s{\bf k}}^{\phantom{\dagger}} \nonumber
\end{eqnarray}
Its calculation is outlined in Appendix \ref{app1}, assuming $\Delta \gg |h|,|\mu|$. As the lowest-order correction to the free energy (\ref{sdF}), (\ref{sdFa}), this diagram produces the following magnetization correction to (\ref{sdZero}):
\begin{eqnarray}\label{sdiM1}
\delta g &=& -n_{i}k_{\textrm{B}}T \; J\eta \, \sinh(\beta h)\tanh(\beta h) \\
\delta m &=& n_{i} \; J\eta \left\lbrack \sinh(\beta h)+\frac{\tanh(\beta h)}{\cosh(\beta h)}\right\rbrack \ . \nonumber
\end{eqnarray}

The quantity $\eta$ plays the same role as the density of states at the Fermi energy in a Kondo metal. It is thermally activated in the low temperature limit $\beta\Delta \gg 1$:
\begin{equation}
\eta = \frac{S_{d}\Gamma\left(\frac{d}{2}\right)(2\beta\Delta)^{d/2}}{(2\pi\beta v)^{d}} \times 2\beta \, e^{-\beta\Delta}\cosh(\beta\mu) \ .
\end{equation}
We see that the Kondo correction to the response of free moments is exponentially sensitive to small magnetic fields, but still thermally activated until the extreme limit $|h|\sim\Delta$.

A decent approximation for $\delta m$ in the $\beta\Delta\ll 1$ limit is given by the above formula with a modified parameter:
\begin{equation}
\eta \approx \left(\frac{\Delta}{2\pi v}\right)^{d} \frac{S_{d}\beta}{\cosh^{2}(\beta\Delta)}
  \begin{cases} C\,(\beta\Delta)^{-d} & ,\quad \beta W \gg 1 \\ C'(W/\Delta)^{d} & ,\quad \beta W \ll 1 \end{cases}  \ .
\end{equation}
The ``constants'' $C$ and $C'$ (dependent on $d$) can be determined by a numerical fit to the exact integral in (\ref{sdiF1b}) at small fields. 

Kondo screening reduces the intrinsic magnetization of free moments in the case of antiferromagnetic coupling $J<0$, since thermally generated particles and holes try to form spin singlets with local moments. This happens in a linear fashion at small fields, i.e. through a renormalization of the impurity magnetic moment. At zero temperature, the Kondo correction to magnetization stays strictly zero until $|h|\gtrsim\Delta$, when it suddenly jumps. Note that free moments at zero temperature immediately saturate in any Zeeman field, and this behavior is not disturbed by the Kondo effect in an insulator.

Specific heat vanishes in zero field at this order of perturbation theory because $\delta g=0$ at $h=0$. We will find a finite thermally activated correction to specific heat only at the second order, where magnetization also acquires its dominant non-activated quantum correction.

Another form of the above result:
\begin{eqnarray}
\delta m &=& \textrm{const}\times n_{i} \, \beta J\left(\frac{mv}{2\pi}\right)^{d} f(\beta\Delta,\beta\mu) \nonumber \\
&& \times \left\lbrack \sinh(\beta h)+\frac{\tanh(\beta h)}{\cosh(\beta h)}\right\rbrack
\end{eqnarray}
provides a more transparent comparison to the second-order quantum correction that was discussed in the introduction; $m=\Delta/v^2$ is the effective mass of low-energy quasiparticles and holes, and $p=mv/2\pi$ is a microscopic energy scale that converts the raw Kondo coupling $J$ to an energy scale $j=J p^d$. It is not hard to see by dimensional analysis that the temperature and field dependence of thermodynamic functions are not qualitatively affected by the precise electron dispersion $\epsilon_{\bf k}$, even in the presence of a spin-orbit coupling. Such details of the electron spectrum can be collected into dimensionless numerical factors and a momentum scale $p$. Using the present model, we can relate $p$ to more objective characteristics of the spectrum:
\begin{equation}
p^{d} \sim \left(\frac{\Delta}{v}\right)^{d}=\left(\frac{\Delta}{W}\right)^{d}\left(\frac{W}{v}\right)^{d}=\rho W\left(\frac{\Delta}{W}\right)^{d}
\end{equation}
such as an energy cut-off $W$ and the average density of electron states $\rho$ that can contribute to Kondo screening (note that $\Lambda \sim W/v$ is a cut-off momentum in the present model, so that $\rho \sim W^{-1} \Lambda^d$).


\subsubsection{Second order corrections: magnetization}\label{secI3b}

Here we analyze magnetization of a Kondo insulator at the second order of perturbation theory. In contrast to the case of a Kondo metal, the dominant part of magnetization in a Kondo insulator appears only at this order -- it originates from virtual particle-hole excitations generated by the Kondo coupling even at $T=0$. Specific heat, however, must remain thermally activated as long as Kramers degeneracy (of local moments) is not lifted or the gap closed.


There are three second order connected vacuum diagrams that appear in the free energy expansion, shown in Fig.\ref{vacdiag}(b)-(d). The diagrams (b) and (c), which contain tadpoles, vanish in zero magnetic field and otherwise are thermally activated. This is formally seen in Appendix \ref{app1}, and easy to understand on physical grounds. A tadpole represents an intra-band process that must be thermally activated because a fully occupied or empty band at zero temperature cannot exhibit spin fluctuations needed for the Kondo interaction.

We will thus start with the most important diagram (d), which is thermally activated in zero field, and finite at $T=0$ when $h\neq 0$. This diagram captures an inter-band process. After a lengthy calculation presented in Appendix \ref{app2-Md}, we find:
\begin{eqnarray}\label{sdiF2d2}
&& F_{2d} = n_{i}V\times\frac{\beta J^{2}}{\Delta}\left(\frac{mv}{2\pi}\right)^{2d}
      \times S_{d}^{2}\,M_{1}^{\phantom{x}}\!\!\left(0,\frac{W}{\Delta}\right) \\
&& \qquad \times \frac{2+3\cosh(\beta i\lambda)\cosh(\beta h)+\cosh(2\beta h)}{\Bigl\lbrack\cosh(\beta i\lambda)
     +\cosh(\beta h)\Bigr\rbrack^{2}}+\mathcal{O}(e^{-\beta\Delta}) \nonumber
\end{eqnarray}
We have introduced the effective mass $m=\Delta/v^2$ of particles and holes, and grouped various factors by meaning. The essential factor that reveals the nature of the second-order perturbative process is $\beta(j^{2}/\Delta)$, where $j=J p^{d}$ is the energy gain of the Kondo coupling between a local moment and a virtual particle-hole pair that intrinsically costs energy $\Delta$. The residual factor of $\beta$ is eliminated in the free energy density $g=g_{0}-(k_{\textrm{B}}T/V)F_{2}$, so the obtained second-order correction is purely a quantum-mechanical shift of the ground state energy. Thermally generated and activated terms have been neglected here. The exact dependence of $F_{2}$ on the cut-off energy scale $W$ in the factor $M_1$ is tied to the high-energy dispersion of electrons and holes -- see Appendix \ref{app2-Md} for details. Using a more realistic non-relativistic dispersion $\epsilon_{\bf k}$ only changes the definition of the momentum scale $p$ that shapes the effective Kondo energy scale $j=J p^{d}$.

The full free energy is contributed also by the diagrams in Fig.\ref{vacdiag}(b,c). With the gained insight, we can easily rule out the diagram (b) as an important contributor at low temperatures because its mobile electron tadpole loops describe only intra-band virtual processes that must be thermally activated or vanish in the absence of magnetic field. In contrast, the diagram (c) contains a particle-hole bubble, which describes inter-band virtual processes. Since particle-hole pairs can be generated by the Kondo interaction even at zero temperature, we ought to explicitly check this diagram -- the calculation presented in Appendix \ref{app2-Mc} shows that this diagram is thermally-activated after all.

In conclusion, \emph{quantum} contributions to the free energy, up to the second order of perturbation theory, come only from (\ref{sdiF2d2}). Using Popov-Fedotov chemical potential $i\lambda=i\pi/2\beta$ and (\ref{ThermoDyn}), (\ref{sdF}) we find the following second order corrections:
\begin{eqnarray}\label{sdiM2}
\delta g &=& -n_{i}k_{\textrm{B}}T_{0}\left\lbrack 2+\frac{1}{\cosh^{2}(\beta h)}\right\rbrack +\mathcal{O}(e^{-\beta\Delta}) \nonumber \\
\delta m &=& -2n_{i}\frac{T_{0}}{T} \frac{\tanh(\beta h)}{\cosh^{2}(\beta h)}+\mathcal{O}(e^{-\beta\Delta}) \ ,
\end{eqnarray}
where we defined a temperature scale $T_0$ by:
\begin{equation}\label{KondoT2}
k_{\textrm{B}}T_{0} = S_{d}^{2}\,M_{1}^{\phantom{x}}\!\!\left(0,\frac{W}{\Delta}\right)\times\frac{J^{2}}{\Delta}\left(\frac{mv}{2\pi}\right)^{2d} \ .
\end{equation}

The intrinsic magnetization of local moments is linearly suppressed at small fields by Kondo screening that involves quantum fluctuations of virtual particle-hole pairs. However, this correction fades away at large fields $h>k_{\textrm{B}}T$ in a thermally activated fashion. Similarly, $\delta m$ fades away both in the limits of zero and infinite temperature when $h$ is kept fixed.

\subsubsection{Second order corrections: specific heat in zero field}\label{secI3c}

The quantum contribution to free energy $\delta g$ in (\ref{sdiM2}) loses temperature dependence in zero field and hence does not provide a correction to specific heat. We must examine the thermally activated terms $\mathcal{O}(e^{-\beta\Delta})$ in order to find a second order correction to specific heat in zero field. To that end, we go back to the diagram $F_{2d}$ shown in Fig.\ref{vacdiag}(d) and specialize to the case $h=0$. The other two second-order diagrams in Fig.\ref{vacdiag}(b,c) have tadpoles and vanish in zero field.

A detailed calculation of the thermally-activated corrections to $F_{2d}$ is presented in Appendix \ref{app2-C}. The main conclusion is that the corresponding specific heat correction behaves as
\begin{equation}\label{dC2a}
\delta c \approx 2Cn_{i}k_{\textrm{B}}\frac{k_{\textrm{B}}T_0}{\Delta}\cosh(\beta\mu)\,
  (\beta\Delta)^{3-\frac{d}{2}}e^{-\beta\Delta}+\mathcal{O}(e^{-2\beta\Delta})
\end{equation}
in the low-temperature $\beta\Delta\gg 1$ limit, and
\begin{equation}\label{dC2b}
\delta c \approx 4n_{i}k_{\textrm{B}}\,\beta^{2}\Delta\, k_{\textrm{B}} T_0 \Bigl\lbrack C_{1}-6C_{2}(\beta\Delta)^{2}+\cdots\Bigr\rbrack \nonumber \ .
\end{equation}
in the high-temperature $\beta\Delta\ll 1$ limit. In the first expression, $C$ is a constant and $T_0$ is a Kondo temperature scale introduced in (\ref{KondoT2}). In the second expression, the constants $C_1>0$ and $C_2$ depend on the ratio $W/\Delta$ between the cut-off energy $W$ and the bandgap $\Delta$.

We see that $\delta c\propto T^{-2}$ exhibits an upturn as the temperature is lowered from the high-temperature limit $\beta\Delta\ll 1$. Therefore, given its thermal activation at lowest temperatures, $\delta c(T)$ must have a peak at intermediate temperatures, in a manner analogous to Schottky anomaly -- but here generated via the Kondo coupling ($T_0 \propto J^2$). The nature of the $\delta c$ upturn evolves and crosses over to a modified temperature dependence in the intermediate regime $\beta\Delta \sim 1$. This is observed in the experiment and described more accurately by (\ref{dC2a}) -- see Fig.\ref{HCapFig}.

\section{Conclusions and discussion}\label{secConclusions}

We calculated magnetization and specific heat in a prototype model of dilute magnetic impurity moments coupled to delocalized electrons of a band insulator. We found that magnetization receives quantum corrections at the second order of perturbation theory due to virtual inter-band particle-hole pairs that partially screen the impurities via Kondo effect. In contrast, specific heat at zero magnetic field is always thermally activated. We worked out the temperature and magnetic field dependence of these quantities, paying special attention to low- and high-temperature regimes. 

Our model is designed to minimalistically describe the physics of isolated magnetic impurities in Kondo insulators and provide physical insight from tractable analytical calculations. This introduces idealizations and approximations which spoil the quantitative accuracy and even the ability to capture some minor qualitative features of realistic Kondo insulators. Perhaps the most dramatic simplification is our treatment of impurities as spin $S=1/2$ moments, whereas in reality Gd impurities in SmB$_6$ have a large moment. Nevertheless, our results agree with thermodynamic experiments \cite{fuhrman2018screened} in crucial ways. They reproduce the essential dependence of magnetization and specific heat on the magnetic impurity concentration, while qualitatively capturing and explaining the effective reduction of impurity moments and the low-temperature specific heat upturn. This requires two channels for Kondo screening, one associated with intrinsic and another with impurity bands. Most importantly, our results for the two-channel Kondo effect in insulators match the relative scaling $\delta m, \delta c \propto n_i J^2$ of magnetization $\delta m$ and specific heat $\delta c$ corrections with the Kondo coupling $J$ and impurity concentration $n_i$ (extracted from different samples \cite{fuhrman2018screened}). The scaling of $\delta m$ and $\delta c$ is mismatched in Kondo metals and different than the measured one \cite{fuhrman2018screened}, so it indirectly reveals the character of low-energy quasiparticles involved in the screening of local moments.

The comparison of our results to thermodynamic experiments paints SmB$_6$ as a true insulator despite some of its metallic-looking features. However, the full spectrum of the observed metallic behaviors in Kondo insulators remains mysterious -- most notably, dHvA quantum oscillations featuring a Lifshitz-Kosevich temperature dependence. This relates to the nature of quasiparticles in Kondo insulators. In the following last section, we discuss the qualitative implications of our findings for the nature of quasiparticles, and invite further studies of a physical mechanism for the contribution of impurity moments to quantum oscillations.

\subsection{Relationship to dHvA quantum oscillations and other probes}

Our results shed light on the low-temperature magnetization and specific heat features in SmB$_6$, which have been viewed as potential evidence of charge-neutral excitations at energy scales below the intrinsic gap. We identified magnetic impurities as the major contributor to these excitations. Our experiment \cite{fuhrman2018screened} specifically scrutinized gadolinium impurities, but one should also note that samarium vacancies can raise the valence of SmB$_6$ toward the magnetic Sm$^{3+}$ valence and thus lead to similar magnetic impurity effects as doped magnetic rare earths. The question is now whether this helps us at all to understand the puzzling dHvA quantum oscillations and other probes.

We pointed out with scaling that magnetization and specific heat behave in a manner more consistent with an insulator than a metal. Our model does not require that the quasiparticles implicated in Kondo screening be charged, but it agrees with the experiment better if we assume that the quasiparticles are gapped. If these quasiparticles are spinons, then the ground state is a gapped spin liquid and it is difficult to explain the observed Lifshitz-Kosevich temperature dependence of bulk quantum oscillations in a wide temperature range \cite{Sebastian2015, Sebastian2018}. So, at least naively, our results are aligned with other experiments \cite{Xiang2013, Fuhrman2014, Xu2016, LuLi2017, Boulanger2018} that rule out the existence of gapless excitations in SmB$_6$ at zero magnetic field -- without contradicting the possibility that a gapless spin liquid could be stabilized at high fields.

Recent quantum oscillation and heat transport experiments \cite{Sebastian2018b, LuLi2018, Matsuda2019} paint YbB$_{12}$, another Kondo insulator, as a more promising candidate for a gapless spin liquid. This material has many similarities to SmB$_6$, but its $f$ electrons are expected to be more localized and correlated than those in SmB$_6$. The Lifshitz-Kosevich temperature dependence of dHvA oscillations \cite{Sebastian2018b, LuLi2018}, which extends to the lowest measured temperatures in YbB$_{12}$ and indicates a Fermi surface, is matched by the evidence of neutral gapless excitations in transport measurements (unlike SmB$_6$). Specific heat also reveals the likely presence of gapless excitations \cite{Matsuda2019}, and features an upturn at low temperatures as in SmB$_6$. It would be interesting to experimentally study the details of this upturn as a function of impurity concentration, and determine whether it can be understood as a result of Kondo screening in a metallic rather than an insulating quasiparticle environment.

Magnetic impurities can contribute to dHvA quantum oscillations. The amount of Kondo screening sensitively depends on the quasiparticle spectrum at broad energy scales -- the bandgap $\Delta$, the energy cut-off $W$ and the density of quasiparticle states $\rho$ all determine the response functions in Kondo insulators, and the analogous facts for Kondo metals have been well-established \cite{Yosida1991}. An external magnetic field that creates Landau orbitals also affects the spectrum at all energy scales. Hence, the Landau quantization of quasiparticle bands should have a significant impact on the amount of Kondo screening. The oscillatory evolution of Landau orbitals with the magnetic field (at any fixed energy) will generate oscillations of the effective screened impurity moment via the Kondo effect. The ensuing oscillating impurity magnetization is a contribution to dHvA effect.

The relative amplitude of these magnetization oscillations expressed as a fraction of the average \emph{impurity} magnetization is independent of the impurity concentration $n_i$, but reflects the strength of the extrinsic Kondo effect according to our model. The \emph{total} magnetization also has an intrinsic Van Vleck component in SmB$_6$, comparable to the impurity component (or larger) only below $\sim 1$\% impurity concentrations in highest saturating magnetic fields of our measurements \cite{fuhrman2018screened}. Therefore, depending on the amount of Kondo screening (which is clearly visible in thermodynamics) and the concentration of all effective magnetic impurities, the relative amplitude of the impurity-based dHvA oscillations could be sizable (this is a prerequisite for having an impact on the observed dHvA effect \cite{Sebastian2015, Sebastian2018}). Since Kondo screening is a quantum effect even in an insulator, thermal activation is not required as in some other prominent interpretations of quantum oscillations \cite{Knolle2015, Knolle2016, Zhang2016, Pal2016, Kumar2017, Fritz2018, Kawakami2019}. In comparison to the spinon Fermi liquid interpretations \cite{Miranda1993, Baskaran2015, Chowdhury2018, Chowdhury2018a}, the relative dHvA oscillation amplitude of impurities is not limited by the density of states in broadened Landau orbitals -- it can be effectively amplified via the new Kondo scale $j$ (which depends on the cut-off).

Further theoretical and experimental studies are needed to obtain reliable estimates of the impurity Kondo temperature and other parameters that enter Eq.\ref{Summary}. Only then it will be possible to calculate the amplitude of quantum oscillations contributed by Kondo impurities and compare its size and temperature dependence with dHvA experiments. Magnetic impurities are clearly important to some probes, and arise both from dopants and vacancies (which are hard to quantify in samples). Therefore, figuring out their impact on dHvA effect could be important for identifying the intrinsic part of the puzzling quantum oscillations in Kondo insulators.

\section{Acknowledgements}

The authors thank Collin Broholm, Tyrel McQueen, Peter Riseborough, Qimiao Si, Brian Skinner, and Debanjan Chowdhury for helpful discussions. This work was supported by the US Department of Energy, office of Basic Energy Sciences, Division of Material Sciences and Engineering under grant DE-FG02-08ER46544. W.T.F. is grateful to the ARCS foundation, Lockheed Martin, and KPMG for the partial support of this work.

\appendix

\section{First-order perturbation theory}\label{app1}

Here we outline the calculation of the first-order Feynman diagram (\ref{sdF1})
\begin{eqnarray}\label{sdF1-app}
&& \!\! F_1 = (-1)^{2} \frac{J}{2\beta}(2\delta_{\alpha\beta'}\delta_{\beta\alpha'}-\delta_{\alpha\alpha'}\delta_{\beta\beta'}) \\
&& \times \sum_{i}\sum_{\omega_n\Omega_n}\sum_s\int\frac{d^{d}k}{(2\pi)^{d}}\;G^{\phantom{x}}_{\alpha\alpha'}(s,{\bf k},\omega_n)D^{ii}_{\beta\beta'}(\Omega_n)
   U_{s{\bf k},s{\bf k}}^{\phantom{\dagger}} \nonumber
\end{eqnarray}
shown in Fig.\ref{vacdiag}(a). Since the electron propagator makes a tadpole loop at the vertex, momentum and band conservation reduces the vertex function (\ref{Usk}) to the trivial form $U_{s{\bf k},s'{\bf k}'}\to 1$. We use the following identities to calculate the sums over repeated spin indices:
\begin{eqnarray}\label{sdIdentities}
\delta_{\alpha\alpha'}(2\delta_{\alpha\beta'}\delta_{\beta\alpha'}-\delta_{\alpha\alpha'}\delta_{\beta\beta'})
  &=& 0 \\
\sigma_{\alpha\alpha'}^{z}(2\delta_{\alpha\beta'}^{\phantom{z}}\delta_{\beta\alpha'}^{\phantom{z}}
      -\delta_{\alpha\alpha'}^{\phantom{z}}\delta_{\beta\beta'}^{\phantom{z}})
  &=& 2\sigma_{\beta\beta'}^{z} \nonumber \\
\sigma_{\alpha\alpha'}^{z}\sigma_{\beta\beta'}^{z}(2\delta_{\alpha\beta'}\delta_{\beta\alpha'}-\delta_{\alpha\alpha'}\delta_{\beta\beta'})
  &=& 2\sigma_{\alpha\alpha'}^{z}\sigma_{\alpha\alpha'}^{z}=4 \nonumber
\end{eqnarray}
The first identity together with (\ref{sdGD}) implies that any diagram with a tadpole vanishes in zero field. Substituting these identities and (\ref{sdGD}) in (\ref{sdF1-app}) gives us:
\begin{eqnarray}
F_1 &=& \frac{Jn_{i}V}{2\beta}\sum_{\sigma\sigma'=\pm 1} \sum_s \int\frac{d^{d}k}{(2\pi)^{d}} \\
  && \; \times \sum_{\omega_n\Omega_n}\frac{\sigma}{i\omega_n-(E_{s{\bf k}}-\mu-h\sigma)}\frac{\sigma'}{i\Omega_n-i\lambda+h\sigma'} \ .\nonumber
\end{eqnarray}
The summation over Matsubara frequencies is carried out by the standard procedure. After a few straight-forward steps we arrive at:
\begin{eqnarray}\label{sdiF1a}
&& F_1 = \frac{J\beta n_{i}V}{2} \tanh^2\left(\frac{\beta h}{2}\right)
  \frac{1-\tanh^{2}\left(\frac{\beta i\lambda}{2}\right)}{1-\tanh^{2}\left(\frac{\beta i\lambda}{2}\right)\tanh^{2}\left(\frac{\beta h}{2}\right)} \nonumber \\
&& ~~ \times \sum_{s}\int\frac{d^{d}k}{(2\pi)^{d}}\frac{1-\tanh^{2}\left(\frac{s\beta\sqrt{\epsilon_{{\bf k}}^{2}+\Delta^{2}}-\beta\mu}{2}\right)}
  {1-\tanh^{2}\left(\frac{s\beta\sqrt{\epsilon_{{\bf k}}^{2}+\Delta^{2}}-\beta\mu}{2}\right)\tanh^{2}\left(\frac{\beta h}{2}\right)} \nonumber
\end{eqnarray}
Using $\epsilon_{{\bf k}}= vk$ allows us to easily introduce a dimensionless energy $\xi=\beta\sqrt{\epsilon_{{\bf k}}^{2}+\Delta^{2}}$ and rewrite momentum integrals as:
\begin{equation}
\int\frac{d^{d}k}{(2\pi)^{d}}
  = \frac{S_{d}}{(2\pi\beta v)^{d}}\int\limits_{\beta\Delta}^{\infty}d\xi\,\xi\Bigl\lbrack\xi^{2}-(\beta\Delta)^{2}\Bigr\rbrack^{\frac{d}{2}-1} \ .
\end{equation} 
After some trigonometric simplifications we arrive at:
\begin{eqnarray}\label{sdiF1b}
F_1 &=& \frac{J\beta n_{i}V}{2}\frac{S_{d}}{(2\pi\beta v)^{d}}\frac{\sinh^2(\beta h)}{\cosh(\beta h)+\cosh(\beta i\lambda)} \\
&& \quad\times \sum_{s}\int\limits_{\beta\Delta}^{\infty}d\xi\,\frac{\xi\Bigl\lbrack\xi^{2}-(\beta\Delta)^{2}\Bigr\rbrack^{\frac{d}{2}-1}}
      {\cosh(\beta h)+\cosh(\xi-s\beta\mu)} \nonumber \\
&& \!\!\!\!\!\!\!\!\!\! \xrightarrow{|h|\ll\Delta} \frac{J\eta}{2}n_{i}V\frac{\sinh^{2}(\beta h)}{\cosh(\beta h)+\cosh(\beta i\lambda)} \ , \nonumber
\end{eqnarray}
The quantity $\eta$ has the units of a density of states and behaves thermally activated in the low temperature limit $\beta\Delta \gg 1$:
\begin{equation}
\eta = \frac{S_{d}\Gamma\left(\frac{d}{2}\right)(2\beta\Delta)^{d/2}}{(2\pi\beta v)^{d}} \times 2\beta \, e^{-\beta\Delta}\cosh(\beta\mu) \ .
\end{equation}
$\eta$ can be similarly approximated in the high-temperature limit $\beta\Delta \ll 1$.

Finally, in order to obtain the magnetization correction written in (\ref{sdiM1}), one has to substitute the Popov-Fedotov chemical potential $i\lambda = i\pi/2\beta$ for localized electrons.

\section{Second-order perturbation theory: magnetization, part 1}\label{app2-Md}

Here we derive the second-order Feynman diagram shown in Fig.\ref{vacdiag}(d):
\begin{eqnarray}\label{sdiF2d}
&& F_{2d} = \frac{(-1)^{2}}{2}\left(\frac{J}{2\beta}\right)^{2}\sum_{ij}\sum_{\omega_{n}^{\phantom{.}}\omega'_{n}}\sum_{\Omega_{n}^{\phantom{.}}}\sum_{ss'}
      \int\frac{d^{d}k}{(2\pi)^{d}}\frac{d^{d}k'}{(2\pi)^{d}} \nonumber \\
&& ~\times e^{i({\bf k}-{\bf k}')({\bf r}_{i}-{\bf r}_{j})} U_{s{\bf k},s'{\bf k}'}^{\phantom{.}} U_{s'{\bf k}',s{\bf k}}^{\phantom{.}} \nonumber \\
&& ~\times (2\delta_{\alpha_{1}^{\phantom{.}}\beta'_{1}}\delta_{\beta_{1}^{\phantom{.}}\alpha'_{1}}
        -\delta_{\alpha_{1}^{\phantom{.}}\alpha'_{1}}\delta_{\beta_{1}^{\phantom{.}}\beta'_{1}})
    (2\delta_{\alpha_{2}^{\phantom{.}}\beta'_{2}}\delta_{\beta_{2}^{\phantom{.}}\alpha'_{2}}
        -\delta_{\alpha_{2}^{\phantom{.}}\alpha'_{2}}\delta_{\beta_{2}^{\phantom{.}}\beta'_{2}}) \nonumber \\
&& ~\times G_{\alpha_1^{\phantom{.}}\alpha'_2}(s,{\bf k},\omega_n) G_{\alpha_2^{\phantom{.}}\alpha'_1}(s',{\bf k}',\omega'_n) \nonumber \\
&& ~\times D_{\beta_1^{\phantom{.}}\beta'_2}^{ij}(\Omega_n) D_{\beta_2^{\phantom{.}}\beta'_1}^{ji}(\Omega_n+\omega_n-\omega'_n) \ .
\end{eqnarray}
The Green's functions of mobile and localized electrons are given by (\ref{sdGD}). The Kronecker symbol $\delta_{\alpha\beta}^{\phantom{.}}$ and the Pauli matrix $\sigma^z_{\alpha\beta}$ in these formulas contract differently their spin indices with the vertices, so we need the means to manage all the terms generated by contractions. To that end, we introduce four new summation variables $\tau_n=\pm1$ to represent the numerators of the four Green's functions in $F_{2d}$:
\begin{equation}
\delta_{\alpha\alpha'}^{\phantom{z}}+\sigma\sigma_{\alpha\alpha'}^{z} =
    \sum_{\tau_n=\pm1} \left\lbrack\frac{1+\tau_n}{2}\delta_{\alpha\alpha'}^{\phantom{z}}+\frac{1-\tau_n}{2}\sigma\sigma_{\alpha\alpha'}^{z}\right\rbrack \nonumber
\end{equation}
in the order $n=1,2,3,4$ of their appearance in (\ref{sdiF2d}). The contraction of spin indices reduces to the following factor that depends on $\tau_n$:
\begin{eqnarray}
S(\tau_n) &=& \frac{1}{2}\left(1+\prod_n\tau_n\right) \biggl\lbrack 2+\frac{3}{2}(\tau_{1}+\tau_{2})(\tau_{3}+\tau_{4}) \\
&& ~~ +\frac{5}{2}(\tau_{1}+\tau_{3})(\tau_{2}+\tau_{4})-\frac{3}{2}(\tau_{1}+\tau_{4})(\tau_{2}+\tau_{3}) \biggr\rbrack \nonumber \ ,
\end{eqnarray}
and we have:
\begin{eqnarray}\label{sdiF2d1}
&& F_{2d} = \frac{n_{i}V}{512}\left(\frac{J}{2\beta}\right)^{2}\sum_{\tau_n\sigma_n} S(\tau_n)
\sum_{\omega_{n}^{\phantom{.}}\omega'_{n}}\sum_{\Omega_{n}^{\phantom{.}}}\sum_{ss'}\int\frac{d^{d}k}{(2\pi)^{d}}\frac{d^{d}k'}{(2\pi)^{d}} \nonumber \\
&& ~~~~~\times \frac{1+\sigma_{1}+(1-\sigma_{1})\tau_{1}}{i\omega_{n}-(E_{s{\bf k}}-\mu-\sigma_{1}h)}\,
          \frac{1+\sigma_{2}+(1-\sigma_{2})\tau_{2}}{i\omega'_{n}-(E_{s'{\bf k}'}-\mu-\sigma_{2}h)}\, \nonumber \\
&& ~~~~~\times \frac{1+\sigma_{3}+(1-\sigma_{3})\tau_{3}}{i\Omega_{n}-i\lambda+\sigma_{3}h}\,
          \frac{1+\sigma_{4}+(1-\sigma_{4})\tau_{4}}{i(\Omega_{n}^{\phantom{.}}+\omega_{n}^{\phantom{.}}-\omega'_{n})-i\lambda+\sigma_{4}h} \nonumber \\
&& ~~~~~\times \Bigl(U_{s{\bf k},s'{\bf k}'}^{\phantom{.}}\Bigr)^{2} \ .
\end{eqnarray}
The impurity site ($i,j$) summation is reduced to the number $N_i=n_i V$ of impurity sites in the volume $V$, and we applied $U_{s{\bf k},s'{\bf k}'} = U_{s'{\bf k}',s{\bf k}}$ according to (\ref{Usk}). This expression is ready for the lengthy but straight-forward summation over Matsubara frequencies:
\begin{eqnarray}\label{sdiF2fsum}
&& \sum_{\omega_{n}^{\phantom{.}}\omega'_{n}}\sum_{\Omega_{n}^{\phantom{.}}}\,
      \frac{1}{i\omega_{n}-(E_{s{\bf k}}-\mu-\sigma_{1}h)}\,\frac{1}{i\omega'_{n}-(E_{s'{\bf k}'}-\mu-\sigma_{2}h)} \nonumber \\
&& \qquad\quad\times \frac{1}{i\Omega_{n}-i\lambda+\sigma_{3}h}\,
      \frac{1}{i(\Omega_{n}^{\phantom{.}}+\omega_{n}^{\phantom{.}}-\omega'_{n})-i\lambda+\sigma_{4}h} \nonumber \\
&& = \frac{\beta^{3}}{8}\biggl\lbrace
      K_1 \frac{2\cosh\left(\frac{\beta(\sigma_{3}-\sigma_{4})h}{2}\right)}
        {\cosh\left(\beta i\lambda-\frac{\beta(\sigma_{3}+\sigma_{4})h}{2}\right)+\cosh\left(\frac{\beta(\sigma_{3}-\sigma_{4})h}{2}\right)} \nonumber \\
&& \qquad~~ - K_2 \frac{(\sigma_{3}-\sigma_{4})\sinh(\beta h)}{\cosh(\beta i\lambda)+\cosh(\beta h)} \biggr\rbrace
\end{eqnarray}
with:
\begin{eqnarray}\label{sdiF2fsum2}
K_1 &=& \frac{\tanh\left(\frac{\beta(E_{s{\bf k}}-\mu-\sigma_{1}h)}{2}\right)-\tanh\left(\frac{E_{s'{\bf k}}-\mu-\sigma_{2}h}{2}\right)}
        {E_{s{\bf k}}-E_{s'{\bf k}'}-(\sigma_{1}-\sigma_{2}+\sigma_{3}-\sigma_{4})h} \\
K_2 &=& \frac{1-\tanh\left(\frac{\beta(E_{s{\bf k}}-\mu-\sigma_{1}h)}{2}\right)\tanh\left(\frac{E_{s'{\bf k}}-\mu-\sigma_{2}h}{2}\right)}
        {E_{s{\bf k}}-E_{s'{\bf k}'}-(\sigma_{1}-\sigma_{2}+\sigma_{3}-\sigma_{4})h} \nonumber \ .
\end{eqnarray}
Next, we will sum over the band-indices $s,s'$. For this, we need to scrutinize the vertex function $U_{ss'}({\bf k},{\bf k}')\equiv U_{s{\bf k},s'{\bf k}'}$ in (\ref{Usk}). One can show that for every ${\bf k}, {\bf k}'$:
\begin{equation}
U_{ss'}({\bf k},{\bf k}')=ss'U_{s's}({\bf k},{\bf k}')\quad,\quad \left(U_{++}\right)^{2}+\left(U_{-+}\right)^{2}=1 \ , \nonumber
\end{equation}
implying:
\begin{eqnarray}\label{Usk2}
U_{ss'} &=& u\delta_{ss'}+s\sqrt{1-u^{2}}(1-\delta_{ss'}) \\
\left(U_{ss'}\right)^{2} &=& u^{2}\delta_{ss'}+(1-u^{2})(1-\delta_{ss'}) \ . \nonumber
\end{eqnarray}
The residual function $u({\bf k}, {\bf k}')$ will be expressed later in a conveniently rescaled form. To sum over $s,s'$ in (\ref{sdiF2d1}), we must combine the vertex function with the $s,s'$-dependent factors (\ref{sdiF2fsum2}) obtained in frequency summations (\ref{sdiF2fsum}):
\begin{eqnarray}
&& Q_1 = \sum_{ss'} K_1 \Bigl\lbrack u^{2}\delta_{ss'}+(1-u^{2})(1-\delta_{ss'})\Bigr\rbrack \\
&& \qquad \!= \frac{4(\xi+\xi')(1-u^{2})}{(\xi+\xi')^{2}-(\sigma_{1}-\sigma_{2}+\sigma_{3}-\sigma_{4})^{2}(\beta h)^{2}}
      +\mathcal{O}(e^{-\beta\Delta}) \nonumber \\[0.2in]
&& Q_2 = \sum_{ss'} K_2 \Bigl\lbrack u^{2}\delta_{ss'}+(1-u^{2})(1-\delta_{ss'})\Bigr\rbrack \nonumber \\
&& \qquad \!= \frac{4\beta h(\sigma_{1}-\sigma_{2}+\sigma_{3}-\sigma_{4})(1-u^{2})}{(\xi+\xi')^{2}-(\sigma_{1}-\sigma_{2}+\sigma_{3}-\sigma_{4})^{2}(\beta h)^{2}}
      +\mathcal{O}(e^{-\beta\Delta}) \nonumber
\end{eqnarray}
We introduced dimensionless energies $\xi=\beta\sqrt{\epsilon_{{\bf k}}^{2}+\Delta^{2}}$ and $\xi'=\beta\sqrt{\epsilon_{{\bf k}'}^{2}+\Delta^{2}}$ to replace momenta ${\bf k}$ and ${\bf k}'$.

For now, we systematically neglect all thermally activated terms by expanding in powers of $1-\tanh(\xi/2)\sim1-\tanh(\xi'/2)\sim e^{-\beta\Delta}$, noting that $\xi,\xi'>\beta\Delta\gg1$. Essentially, the intra-band Kondo scattering (proportional to $u^{2}$) is thermally activated, but inter-band Kondo scattering (proportional to $1-u^{2}$) is not. Also, $\cosh(\beta\mu)$ is negligible next to $\cosh(\xi)$ or $\cosh(\xi')$ when $\Delta\gg|\mu|$. 

Now, we are ready to integrate out momenta. We will benefit from changing the momentum integration variables into $x,y$, where $\beta\Delta(1+x)=(\xi+\xi')/2$ and $2\beta\Delta y=\xi-\xi'$. The integrals expressed in terms of $x,y$ will be temperature-independent, and will isolate well their dependence on magnetic field $h$. Their ultra-violet divergence will be controlled by the effective bandwidth $W$. We have:
\begin{eqnarray}
&& \int\frac{d^{d}k}{(2\pi)^{d}}\frac{d^{d}k'}{(2\pi)^{d}}\,Q_{1/2} = \frac{8S_{d}^{2}}{\beta\Delta}\left(\frac{mv}{2\pi}\right)^{2d} \\
&& \qquad \times M_{1/2} \left((\sigma_{1}-\sigma_{2}+\sigma_{3}-\sigma_{4})\frac{h}{2\Delta},\frac{W}{\Delta}\right)
   +\mathcal{O}(e^{-\beta\Delta}) \nonumber \ ,
\end{eqnarray}
where the functions $M_1$ and $M_2$ are dimensionless integrals:
\begin{eqnarray}
&& M_{i}(\chi,w) = \int\limits_{0}^{w}d x\int\limits_{0}^{x}dy\,
    \Bigl\lbrace(x^2-y^2)\left\lbrack(x+2)^2-y^2\right\rbrack\Bigr\rbrace^{\frac{d}{2}-1} \nonumber \\
&& \qquad\times \frac{(x+1)^{2}-y^{2}}{(x+1)^{2}-\chi^{2}}(1-u^{2}) \Bigl\lbrack(x+1)\delta_{i,1}+\chi\delta_{i,2}\Bigr\rbrack
\end{eqnarray}
with:
\begin{eqnarray}
&& u(x,y) = \frac{1}{4}\,\frac{1}{\sqrt{(x+1)^{2}-y^{2}}} \\
&& ~\times \frac{4\!+\!\left(\sqrt{x+y}-\sqrt{x+y+2}\right)^{2}\left(\sqrt{x-y}-\sqrt{x-y+2}\right)^{2}}
       {\left(\sqrt{x+y}-\sqrt{x+y+2}\right)\left(\sqrt{x-y}-\sqrt{x-y+2}\right)} \nonumber
\end{eqnarray}
We will not need the values of $M_i$, except:
\begin{equation}
M_1(0,w) \xrightarrow{w\gg1} \frac{w^3}{9} \qquad \textrm{in}\;d=3 \ .
\end{equation}
Putting everything together into (\ref{sdiF2d1}) and writing compactly $\chi=(\sigma_{1}-\sigma_{2}+\sigma_{3}-\sigma_{4})h/2\Delta$, we obtain:
\begin{eqnarray}
&& F_{2d} = \frac{n_{i}V\beta^4}{512}\left(\frac{J}{2\beta}\right)^{\!\!2}
    \!\!\sum_{\tau_n\sigma_n}S(\tau_n)\prod_n\Bigl\lbrack1+\sigma_{n}+(1-\sigma_{n})\tau_{n}\Bigr\rbrack \nonumber \\
&& ~~ \times\frac{S_{d}^{2}}{\beta\Delta}\left(\frac{mv}{2\pi}\right)^{2d} \Biggl\lbrack
         -\frac{(\sigma_{3}-\sigma_{4})\sinh(\beta h)\,M_{2}\left(\chi,\frac{W}{\Delta}\right)}{\cosh(\beta i\lambda)+\cosh(\beta h)} \nonumber \\
&& ~~~~ + \frac{2\cosh\left(\frac{\beta(\tau_{3}-\tau_{4})h}{2}\right) M_{1}\left(\chi,\frac{W}{\Delta}\right)}
         {\cosh\left(\beta i\lambda-\frac{\beta(\tau_{3}+\tau_{4})h}{2}\right)+\cosh\left(\frac{\beta(\tau_{3}-\tau_{4})h}{2}\right)} \Biggr\rbrack
         + \cdots \nonumber
\end{eqnarray}
up to the thermally activated terms ($\cdots$). Finally, we sum over $\sigma_n$ and $\tau_n$ to obtain a relatively simple expression written in the main text (\ref{sdiF2d2}):
\begin{eqnarray}\label{sdiF2d2-app}
&& F_{2d} = n_{i}V\times\frac{\beta J^{2}}{\Delta}\left(\frac{mv}{2\pi}\right)^{2d}
      \times S_{d}^{2}\,M_{1}^{\phantom{x}}\!\!\left(0,\frac{W}{\Delta}\right) \\
&& \qquad \times \frac{2+3\cosh(\beta i\lambda)\cosh(\beta h)+\cosh(2\beta h)}{\Bigl\lbrack\cosh(\beta i\lambda)
     +\cosh(\beta h)\Bigr\rbrack^{2}}+\mathcal{O}(e^{-\beta\Delta}) \nonumber
\end{eqnarray}

\section{Second-order perturbation theory: magnetization, part 2}\label{app2-Mc}

Here we show that the second-order Feynman diagram in Fig.\ref{vacdiag}(c) is thermally activated. It is immediately evident that this diagram vanishes in zero field due to its tadpoles. The initial formula for this diagram is:
\begin{eqnarray}\label{sdiF2c1}
&& F_{2c} = \frac{(-1)^{3}}{2}\left(\frac{J}{2\beta}\right)^{2}\sum_{ij}\sum_{\Omega_{n}^{\phantom{.}}\Omega'_n}\sum_{\omega_{n}^{\phantom{.}}}\sum_{ss'}
    \int\frac{d^{d}k}{(2\pi)^{d}}\frac{d^{d}k'}{(2\pi)^{d}} \nonumber \\
&& ~\times e^{i({\bf k}-{\bf k}')({\bf r}_{i}-{\bf r}_{j})} U_{s{\bf k},s'{\bf k}'}^{\phantom{.}}U_{s'{\bf k}',s{\bf k}}^{\phantom{.}} \nonumber \\
&& ~\times (2\delta_{\alpha_{1}^{\phantom{.}}\beta'_{1}}\delta_{\beta_{1}^{\phantom{.}}\alpha'_{1}}
             -\delta_{\alpha_{1}^{\phantom{.}}\alpha'_{1}}\delta_{\beta_{1}^{\phantom{.}}\beta'_{1}})
            (2\delta_{\alpha_{2}^{\phantom{.}}\beta'_{2}}\delta_{\beta_{2}^{\phantom{.}}\alpha'_{2}}
             -\delta_{\alpha_{2}^{\phantom{.}}\alpha'_{2}}\delta_{\beta_{2}^{\phantom{.}}\beta'_{2}}) \nonumber \\
&& ~\times G_{\alpha_1^{\phantom{.}}\alpha'_2}(s,{\bf k},\omega_n) G_{\alpha_2^{\phantom{.}}\alpha'_1}(s',{\bf k}',\omega_n) \nonumber \\
&& ~\times D_{\beta_1^{\phantom{.}}\beta'_1}^{ii}(\Omega_n) D_{\beta_2^{\phantom{.}}\beta'_2}^{jj}(\Omega'_n) \ .
\end{eqnarray}
We will first contract all spin indices. After some manipulations, we arrive at:
\begin{eqnarray}
&& F_{2c} = -\frac{1}{2}\left(\frac{J}{2\beta}\right)^{2}\sum_{ij}\sum_{\omega_{n}^{\phantom{.}}}\sum_{ss'}
    \int\frac{d^{d}k}{(2\pi)^{d}}\frac{d^{d}k'}{(2\pi)^{d}} \nonumber \\
&& ~\times e^{i({\bf k}-{\bf k}')({\bf r}_{i}-{\bf r}_{j})} U_{s{\bf k},s'{\bf k}'}^{\phantom{.}}U_{s'{\bf k}',s{\bf k}}^{\phantom{.}} \nonumber \\
&& ~\times \!\!\left( \sum_\sigma \frac{1}{i\omega_{n}-(E_{s{\bf k}}-\mu-h\sigma)}\,\frac{1}{i\omega_{n}-(E_{s'{\bf k}'}-\mu-h\sigma)} \right) \nonumber \\
&& ~\times \!\!\left( \sum_{\Omega_{n}}\sum_{\sigma=\pm 1} \frac{\sigma}{i\Omega_{n}-i\lambda+h\sigma} \right)^2 \ .
\end{eqnarray}
Summing up Matsubara frequencies yields:
\begin{eqnarray}\label{sdiF2c2}
&& F_{2c} = \left(\frac{J}{2\beta}\right)^{2}\frac{\beta^{3}}{16}\sum_{ss'}\int\frac{d^{d}k}{(2\pi)^{d}}\frac{d^{d}k'}{(2\pi)^{d}}\;
      \sum_{ij}e^{i({\bf k}-{\bf k}')({\bf r}_{i}-{\bf r}_{j})} \nonumber \\
&& \qquad\times \frac{\Bigl(U_{s{\bf k},s'{\bf k}'}^{\phantom{.}}\Bigr)^{2}}{E_{s{\bf k}}-E_{s'{\bf k}'}}
      \left\lbrack \frac{2\sinh(\beta h)}{\cosh(\beta i\lambda)+\cosh(\beta h)}\right\rbrack ^{2} \nonumber \\
&& \qquad\times \biggl\lbrack\frac{2\sinh(\beta(E_{s{\bf k}}-\mu))}{\cosh(\beta(E_{s{\bf k}}-\mu))+\cosh(\beta h)} \nonumber \\
&& \qquad\qquad -\frac{2\sinh(\beta(E_{s'{\bf k}'}-\mu))}{\cosh(\beta(E_{s'{\bf k}'}-\mu))+\cosh(\beta h)}\biggr\rbrack \ .
\end{eqnarray}
This diagram involves a non-trivial summation over the impurity positions ${\bf r}_i$. Diagrams of this kind can generate RKKY-type interactions between proximate local moments. The summation over ${\bf r}_{i}$ and ${\bf r}_{j}$ is equivalent to the summation over $\bar{{\bf r}}=\frac{1}{2}({\bf r}_{i}+{\bf r}_{j})$
and $\delta{\bf r}={\bf r}_{i}-{\bf r}_{j}$. In a particular realization of impurity disorder, the impurity sites ${\bf r}_{i}$ are randomly scattered with some average spatial separation $a$. However, the distribution of $\delta{\bf r}$ is expected to significantly and broadly extend below $|\delta{\bf r}|<a$ because there are many neighboring impurities separated by arbitrarily short distances on the scale of the entire sample. Assuming that impurity locations are not mutually correlated, the translationally invariant distribution of $\delta{\bf r}$ allows us to treat it as a continuous uniform random variable (it gets averaged over the entire system volume). Therefore, we may approximate:
\begin{eqnarray}
&& \sum_{ij}e^{i({\bf k}-{\bf k}')({\bf r}_{i}-{\bf r}_{j})}\approx\sum_{\bar{{\bf r}}}\frac{1}{a^{d}}\int d^{d}\delta r\,e^{i({\bf k}-{\bf k}')\delta{\bf r}} \\
&& \qquad\qquad = \frac{N_{i}}{a^{d}}(2\pi)^{d}\delta({\bf k}-{\bf k}')=\frac{n_{i}V}{a^{d}}(2\pi)^{d}\delta({\bf k}-{\bf k}') \nonumber \\
&& \qquad\qquad = n_{i}^{2}V\times(2\pi)^{d}\delta({\bf k}-{\bf k}') \nonumber
\end{eqnarray}
where $N_{i}=n_{i}V\sim V/a^{d}$ is the total number of impurities.

Once ${\bf k}'$ becomes equal to ${\bf k}$, the vertex function $U_{s{\bf k},s'{\bf k}'}\to\delta_{ss'}$ becomes trivial and forces the two electron propagators to carry the same band index. However, we must take the limit ${\bf k}'\to{\bf k}$ and $s'=s$ carefully because the integrand of (\ref{sdiF2c2}) becomes singular:
\begin{eqnarray}
&& \lim_{{\bf k}'\to{\bf k}}\frac{\frac{2\sinh(\beta(E_{s{\bf k}}-\mu))}{\cosh(\beta(E_{s{\bf k}}-\mu))+\cosh(\beta h)}-\frac{2\sinh(\beta(E_{s{\bf k}'}-\mu))}{\cosh(\beta(E_{s{\bf k}'}-\mu))+\cosh(\beta h)}}{E_{s{\bf k}}-E_{s{\bf k}'}} \nonumber \\
&& \qquad = \frac{\partial}{\partial E} \frac{2\sinh(\beta(E-\mu))}{\cosh(\beta(E-\mu))+\cosh(\beta h)} \biggr\vert_{E=E_{s{\bf k}}} \nonumber \\
&& \qquad = 2\beta\frac{1+\cosh(\beta(E_{s{\bf k}}-\mu))\cosh(\beta h)}{\lbrack\cosh(\beta(E_{s{\bf k}}-\mu))+\cosh(\beta h)\rbrack^{2}} \ .
\end{eqnarray}
Resolving the singularity this way and then integrating disorder is physically motivated because the distribution of $\delta{\bf r}$ is infra-red cut off by the system size, just like the quantized values of momentum ${\bf k}$. We should obtain some thermodynamic effect from very small $|{\bf k}'-{\bf k}|$, as captured here. We now have:
\begin{eqnarray}
F_{2c} &=& \frac{n_{i}^{2}V\beta^{4}}{8} \left(\frac{J}{2\beta}\right)^{2}
    \left\lbrack \frac{2\sinh(\beta h)}{\cosh(\beta i\lambda)+\cosh(\beta h)}\right\rbrack ^{2} \nonumber \\
&& \times \sum_{s}\int\frac{d^{d}k}{(2\pi)^{d}}
    \frac{1+\cosh(\beta(E_{s{\bf k}}-\mu))\cosh(\beta h)}{\lbrack\cosh(\beta(E_{s{\bf k}}-\mu))+\cosh(\beta h)\rbrack^{2}} \nonumber \\
&=& \mathcal{O}(e^{-\beta\Delta}) \ .
\end{eqnarray}
There is no need to calculate any further because this diagram is clearly thermally activated: $\beta|E_{s{\bf k}}|\ge\beta\Delta\gg1$ makes the denominator with $\cosh(\beta(E_{s{\bf k}}-\mu))$ exponentially large at any magnetic field $|h|\ll\Delta$. Also, physically, no particle-hole processes remain after impurity-position summation.

\section{Second-order perturbation theory: specific heat in zero field}\label{app2-C}

Here we calculate the thermally-activated corrections to the diagram shown in Fig.\ref{vacdiag}(d), specializing to the zero magnetic field $h=0$. The calculation in $h=0$ is considerably simpler because the Green's functions (\ref{sdGD}) reduce to:
\begin{eqnarray}\label{sdGDh0}
G_{\alpha\alpha'}(s,{\bf k},\omega_{n}) &=& \frac{\delta_{\alpha\alpha'}}{i\omega_{n}-(E_{s{\bf k}}-\mu)} \nonumber \\
D^{ij}_{\beta\beta'}(\Omega_{n}) &=& \frac{\delta_{\beta\beta'}\delta_{ij}}{i\Omega_{n}-i\lambda} \ .
\end{eqnarray}
Substituting in (\ref{sdiF2d}) and using the first spin-index identity of (\ref{sdIdentities}) quickly gives us:
\begin{eqnarray}
&& F_{2d} = \frac{3}{2}\left(\frac{J}{\beta}\right)^{2}n_{i}V\sum_{\omega_{n}^{\phantom{.}}\omega'_{n}}\sum_{\Omega_{n}^{\phantom{.}}}\sum_{ss'}
  \int\frac{d^{d}k}{(2\pi)^{d}}\frac{d^{d}k'}{(2\pi)^{d}} \\
&& \quad\times U_{s{\bf k},s'{\bf k}'}U_{s'{\bf k}',s{\bf k}} \,
      \frac{1}{i\omega_{n}^{\phantom{.}}-(E_{s{\bf k}}-\mu)}\,\frac{1}{i\omega'_{n}-(E_{s'{\bf k}'}-\mu)} \nonumber \\
&& \qquad\qquad\qquad\quad\;\,\times\frac{1}{i(\Omega_{n}^{\phantom{.}}-\lambda)}\,\frac{1}{i(\Omega_{n}^{\phantom{.}}+\omega_{n}^{\phantom{.}}-\omega'_{n}-\lambda)}
   \ .  \nonumber
\end{eqnarray}
Summing up the Matsubara frequencies results with an expression analogous to (\ref{sdiF2fsum}) and (\ref{sdiF2fsum2}):
\begin{equation}\label{sdiF2d3}
F_{2d} = \frac{3}{16}\frac{\beta J^{2}n_{i}V}{\cosh^{2}\left(\frac{\beta i\lambda}{2}\right)}\sum_{ss'}\int\frac{d^{d}k}{(2\pi)^{d}}\frac{d^{d}k'}{(2\pi)^{d}}
     \Bigl(U_{s{\bf k},s'{\bf k}'}\Bigr)^2 K_1^{\phantom{.}}
\end{equation}
with:
\begin{equation}
K_1 = \frac{\tanh\left(\frac{\beta(E_{s{\bf k}}-\mu)}{2}\right)-\tanh\left(\frac{\beta(E_{s'{\bf k}'}-\mu)}{2}\right)}
        {E_{s{\bf k}}-E_{s'{\bf k}'}} \ .
\end{equation}
The following steps are also similar to the analysis of Appendix \ref{app2-Md}, but depart from it by scrutinizing the thermally activated terms. Expressing the vertex function as (\ref{Usk2}), we carry out the summation over band indices exactly in the above formula:
\begin{eqnarray}\label{Q1}
&& Q_1 = \sum_{ss'} K_1 \Bigl\lbrack u^{2}\delta_{ss'}+(1-u^{2})(1-\delta_{ss'})\Bigr\rbrack \\
&& ~ =\frac{u^{2}}{\xi-\xi'}\left\lbrack \frac{2\sinh(\xi)}{\cosh(\beta\mu)+\cosh(\xi)}-\frac{2\sinh(\xi')}{\cosh(\beta\mu)+\cosh(\xi')}\right\rbrack \nonumber \\
&& ~~ +\frac{1-u^{2}}{\xi+\xi'}\left\lbrack \frac{2\sinh(\xi)}{\cosh(\beta\mu)+\cosh(\xi)}+\frac{2\sinh(\xi')}{\cosh(\beta\mu)+\cosh(\xi')}\right\rbrack \nonumber
\end{eqnarray}
where $\xi=\beta\sqrt{\epsilon_{{\bf k}}^{2}+\Delta^{2}}$ and $\xi'=\beta\sqrt{\epsilon_{{\bf k}'}^{2}+\Delta^{2}}$. From this point on, we will separately consider the low-temperature $\beta\Delta\gg 1$ and high-temperature $\beta\Delta\ll 1$ limits -- both are accessible in perturbation theory when the energy scale $J^2(mv/2\pi)^{2d}/\Delta$ is small enough.

In the low-temperature regime, we can approximate $\sinh(\xi) \approx \cosh(\xi) \approx \frac{1}{2}e^\xi$ because $\xi>\beta\Delta\gg 1$. This leads to:
\begin{eqnarray}
Q_1^{\phantom{0}} &=& Q_{1}^{(0)}-4\cosh(\beta\mu)\Biggl\lbrack u^{2}\frac{e^{-\xi}-e^{-\xi'}}{\xi-\xi'} \\
&& \qquad\qquad +(1-u^{2})\frac{e^{-\xi}+e^{-\xi'}}{\xi+\xi'}\Biggr\rbrack +\mathcal{O}(e^{-2\beta\Delta}) \ , \nonumber
\end{eqnarray}
where $ Q_{1}^{(0)}$ is the non-thermally activated part that we dealt with in Appendix \ref{app2-Md}. Substituting in (\ref{sdiF2d3}) yields:
\begin{eqnarray}
F_{2d}^{\phantom{0}} &=& F_{2d}^{(0)}+\frac{n_{i}V}{\cosh^{2}\left(\frac{\beta i\lambda}{2}\right)}\frac{\beta J^{2}}{\Delta}\left(\frac{mv}{2\pi}\right)^{2d}
  e^{-\beta\Delta}\cosh(\beta\mu) \nonumber \\
&& \qquad \times M'\left(\frac{W}{\Delta},\beta\Delta\right)+\mathcal{O}(e^{-2\beta\Delta}) \ ,
\end{eqnarray}
where:
\begin{eqnarray}
&& M'\left(w,\beta\Delta\right) = \frac{3S_{d}^{2}}{2}\int\limits_{0}^{w}dx\int\limits_{0}^{x}dy\,
    \Bigl((x+1)^{2}-y^{2}\Bigr) \\
&& ~~~ \times \Bigl((x+y)(x+y+2)\Bigr)^{\frac{d}{2}-1}
       \Bigl((x-y)(x-y+2)\Bigr)^{\frac{d}{2}-1} \nonumber \\
&& ~~~ \times e^{-\beta\Delta x}\left\lbrack u^{2}\frac{\sinh(\beta\Delta y)}{y}
        -(1-u^{2})\frac{\cosh(\beta\Delta y)}{x+1}\right\rbrack \nonumber \nonumber
\end{eqnarray}
is expressed using the dimensionless variables $x,y$ defined by $\beta\Delta(1+x)=(\xi+\xi')/2$ and $2\beta\Delta y=\xi-\xi'$, which we introduced in the previous section.  The quantum term $F_{2d}^{(0)}$ is given by (\ref{sdiF2d2}) in $h=0$ and does not contribute to specific heat. We must understand the temperature dependence of $M'$. Crudely, the divergent part of the integral involving $x\to w$ near the cut-off is dominated by $y\approx x$, because for $y<x$ the factors $e^{-\beta\Delta(x-y)}$ that approximate the $\sinh,\cosh$ factors become exponentially suppressed. Thus, we can substitute $y\to x$ almost everywhere in the integral except inside $\sinh,\cosh$ and one factor of $x-y$. The resulting approximation is:
\begin{eqnarray}
&& M'\left(w,\beta\Delta\right) \approx \frac{3S_{d}^{2}}{4}8^{\frac{d}{2}-1}\int\limits_{0}^{w}dx\,e^{-\beta\Delta x}\Bigl(2x+1\Bigr) \\
&& \qquad \times \Bigl(x(x+1)\Bigr)^{\frac{d}{2}-1}\left(\frac{u^{2}}{x}-\frac{1-u^{2}}{x+1}\right)\Biggr\vert_{y=x} \times I_y \nonumber
\end{eqnarray}
with:
\begin{eqnarray}
&& I_y = \int\limits_{0}^{x}dy\,\Bigl((x-y)\Bigr)^{\frac{d}{2}-1}e^{\beta\Delta y} = \frac{e^{\beta\Delta x}}{(\beta\Delta)^{\frac{d}{2}}}
  \int\limits_{0}^{\beta\Delta x}dt\,t^{\frac{d}{2}-1}e^{-t} \nonumber \\
&& \qquad\qquad \xrightarrow{\beta\Delta x\gg1} \frac{e^{\beta\Delta x}}{(\beta\Delta)^{\frac{d}{2}}} \Gamma\left(\frac{d}{2}\right) \ .
\end{eqnarray}
The remaining integration over $x$ is temperature-independent, so we conclude:
\begin{equation}
F_{2d}^{\phantom{x}}\approx F_{2d}^{(0)}+\frac{C\,n_{i}V\,\beta k_{\textrm{B}}T_0}{\cosh^{2}\left(\frac{\beta i\lambda}{2}\right)} \,
  \cosh(\beta\mu)\frac{e^{-\beta\Delta}}{(\beta\Delta)^{\frac{d}{2}}}+\mathcal{O}(e^{-2\beta\Delta}) \ ,
\end{equation}
where $C$ is a constant and $T_0$ is a Kondo temperature scale introduced in (\ref{KondoT2}). It follows that ($i\lambda=i\pi/2\beta$):
\begin{eqnarray}
\delta g &=& \delta g^{(0)}-2C n_{i}k_{\textrm{B}}T_0\,\cosh(\beta\mu)\frac{e^{-\beta\Delta}}{(\beta\Delta)^{\frac{d}{2}}}
  +\mathcal{O}(e^{-2\beta\Delta}) \nonumber \\
\delta c &=& 2Cn_{i}k_{\textrm{B}}\frac{k_{\textrm{B}}T_0}{\Delta}\cosh(\beta\mu)\,
  (\beta\Delta)^{3-\frac{d}{2}}e^{-\beta\Delta}+\mathcal{O}(e^{-2\beta\Delta}) \nonumber
\end{eqnarray}
in the low-temperature $\beta\Delta\gg 1$ limit.

Next, we analyze the high-temperature limit. For simplicity, we will take $\cosh(\beta\mu)\approx 1$ and then expand (\ref{Q1}) in powers of $\beta\Delta\ll 1$:
\begin{eqnarray}
Q_1 &\approx& 2u^{2}\,\frac{\tanh\left(\frac{\xi}{2}\right)-\tanh\left(\frac{\xi'}{2}\right)}{\xi-\xi'} \\
&& +2(1-u^{2})\,\frac{\tanh\left(\frac{\xi}{2}\right)+\tanh\left(\frac{\xi'}{2}\right)}{\xi+\xi'} \nonumber \\
&=& 1-\frac{(\beta\Delta)^{2}}{12}\Bigl\lbrack(x+1)^{2}(2u^{2}+1)+y^{2}(3-2u^{2})\Bigr\rbrack+\cdots \nonumber
\end{eqnarray}
Since $x$ is integrated out up to $w=W/\Delta$, where $W$ is the bandwidth, this expansion is actually in powers of $\beta W$ -- which we assume to be small. The ensuing condition $T\gg \Delta$ is the only path available in the present insulating model toward a specific heat that exhibits a Schottky-like upturn when temperature is reduced over a certain range, as seen in the experiments on SmB$_6$. This forces us to interpret carefully the meaning of the gap $\Delta$, given that the upturn is seen down to millikelvin temperatures. An interpretation of our results and experiments is discussed in the introduction; here, we simply finish presenting the derivations. Substituting $Q_1$ into $F_{2d}$ yields:
\begin{equation}
F_{2d} \approx \frac{n_{i}V\,\beta^{2}\Delta\, k_{\textrm{B}} T_0}{\cosh^{2}\left(\frac{\beta i\lambda}{2}\right)}
  \Bigl\lbrack C_{1}-C_{2}(\beta\Delta)^{2}+\cdots\Bigr\rbrack \ ,
\end{equation}
and then ($i\lambda=i\pi/2\beta$):
\begin{eqnarray}
\delta g &\approx& -2n_{i}\,\beta \Delta\, k_{\textrm{B}} T_0 \Bigl\lbrack C_{1}-C_{2}(\beta\Delta)^{2}+\cdots\Bigr\rbrack \nonumber \\
\delta c &\approx& 4n_{i}k_{\textrm{B}}\,\beta^{2}\Delta\, k_{\textrm{B}} T_0 \Bigl\lbrack C_{1}-6C_{2}(\beta\Delta)^{2}+\cdots\Bigr\rbrack \nonumber \ .
\end{eqnarray}
The constants $C_1>0$ and $C_2$ depend on $W/\Delta$.

\newpage


%

\end{document}